\documentclass[aps,prx,twocolumn,floatfix,longbibliography,nofootinbib,tbtags,10pt]{revtex4-2}
\usepackage[utf8]{inputenc}
\usepackage{natbib}
\usepackage{graphicx}
\usepackage{xcolor}
\usepackage{mathtools}
\usepackage{amsmath}
\usepackage[colorlinks, linkcolor=blue]{hyperref}
\hypersetup{colorlinks,allcolors=black}
\usepackage{amssymb}
\usepackage{gensymb}
\usepackage{float}
\usepackage{txfonts}
\usepackage{tabularx,graphicx}
\usepackage{epstopdf}
\usepackage{latexsym}
\usepackage{color, colortbl}
\usepackage{psfrag}
\usepackage{bbm}
\usepackage{bm}
\usepackage{titlesec}
\usepackage{dsfont}
\usepackage{feynmp}
\usepackage{slashed}
\usepackage{multirow}
\usepackage[normalem]{ulem}
\renewcommand{\vec}[1]{\boldsymbol{#1}}
\def \nn {\nonumber}

\def \k {{\vec k}}
\def \p {{\vec p}}

\def \q {{\vec q}}

\def \Q{{\cal Q}}

\def \l {\ell}

\def \G {{\vec{G}}}

\def \l {\ell}

\def \beq {\begin{eqnarray}}
\def \eeq {\end{eqnarray}}
\def \hi {H_{\tn{int}}}
\def \hk {H_{\tn{kin}}}
\def \tn {\textnormal}

\def \hk {H_{\tn{kin}}}
\def \hi {H_{\tn{int}}}
\def \kxx {K_{\tn{xx}}^{\tn{eff}}}
\def \l {\ell}

\begin{document}
\title{Low-energy optical sum-rule in moir\'e graphene}
\author{J.F. Mendez-Valderrama}
\author{Dan Mao}
\author{Debanjan Chowdhury}
\affiliation{Department of Physics, Cornell University, Ithaca, New York 14853, USA.}

\begin{abstract}
Few layers of graphene at small twist-angles have emerged as a fascinating platform for studying the problem of strong interactions in regimes with a nearly quenched single-particle kinetic energy and non-trivial band topology. 
Starting from the strong-coupling limit of twisted bilayer graphene with a vanishing single-electron bandwidth and interlayer-tunneling between the same sublattice sites, we present an {\it exact} analytical theory of the Coulomb interaction-induced low-energy optical spectral weight at all {\it integer} fillings. In this limit, while the interaction-induced single-particle dispersion is finite, the optical spectral weight vanishes identically at integer fillings. We study corrections to the optical spectral weight by systematically including the effects of experimentally relevant strain-induced renormalization of the single-electron bandwidth and interlayer tunnelings between the same sublattice sites. 
Given the relationship between the optical spectral weight and the diamagnetic response that controls superconducting $T_c$, our results highlight the relative importance of specific parent insulating phases in enhancing the tendency towards superconductivity when doped away from integer fillings. 

\end{abstract}

\maketitle

{\bf Introduction.-} Sum-rules in many-particle quantum mechanics impose universal constraints on physical observables \cite{sumrule,TR,kuhn} that might otherwise be difficult to evaluate from first principles. They relate specific dynamical correlation functions, involving matrix elements of local operators between many-body eigenstates, to physical properties of the system, such as the electron density, mass, and so on. 
In correlated electron systems, sum-rules are routinely used for extracting useful information about experimentally measured correlation functions \cite{pines2018elementary}. 
A famous example --- the optical sum-rule  (or the $f-$sum-rule) --- relates the integral of the longitudinal optical conductivity over all frequencies to the diamagnetic response summed over the entire electronic bandwidth \cite{kubo,hazra2019bounds}. In this traditional setting, all electronic solids have a total optical spectral weight that is on the order of a few electron-volts; the total optical spectral weight is {\it not} a low-energy property, and therefore likely not directly relevant for the physical properties of interest to us. However, a low-energy formulation of an optical sum-rule going beyond the traditional setting, that focuses only on a subset of the relevant electronic bands \cite{Bistritzer2011}, has important experimental implications \cite{DMDC1}.  
In particular, it can serve as a useful upper bound on the maximum superconducting transition temperature in two dimensions  \cite{MR21,DMDC1,DMDC2}.

\begin{figure}[pth!]
\centering
\includegraphics[width=0.85\linewidth]{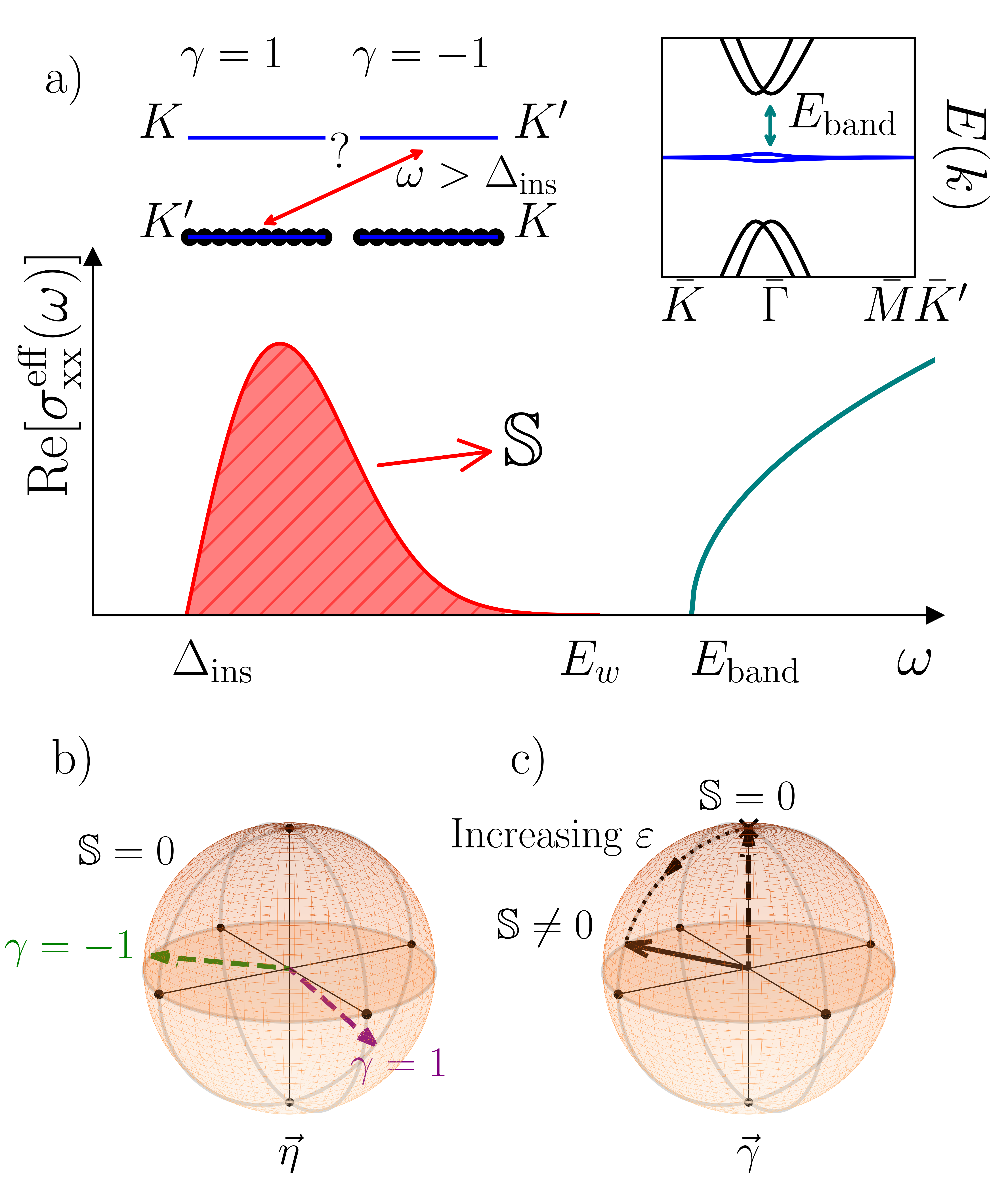}
\caption{(a) Low-energy optical spectral weight, $\mathbb{S}$ (red hatched region), at an integer filling of the isolated, nearly flat bands, due to a finite optical absorption for $\omega>\Delta_{\rm{ins}}$. Top left panel: The many-body state at integer filling with opposite Chern numbers $\gamma=\pm1$ (and the two valleys, $K,~K'$) occupied. Top right panel:  Non-interacting continuum model bandstructure \cite{Bistritzer2011} for TBG at $\theta=1.06$, with tunneling ratio, $w_0/w_1=0.75$. (b) The valley pseudospin, $\vec{\eta}$, tied to the momentum-independent spin-polarized many-body ground states at $\nu = -2$ with total $\gamma = 0$. For the entire family of states, $\mathbb{S}=0$. (c) Representation of the components of the many-body ground states on the Chern pseudospin Bloch sphere ($\vec{\gamma}$). The spectral weight $\mathbb{S}=0$ when the (momentum-independent) density matrix commutes with $\gamma_z$ (at `$\times$'). With inter-Chern coherence, $\mathbb{S}$ acquires a finite value which can be tuned by strain, $\varepsilon$ (e.g. along dashed trajectory).  }
\label{fig:schematic}
\end{figure}

An experimentally relevant example arises in the context of moir\'e ``flat" band materials, such as  twisted bilayer graphene (TBG). While the traditional $f-$sum-rule must necessarily include all intra and inter-band contributions from the entire bandwidth upto the $O(\tn{eV})$ scale, computing the low-energy optical spectral weight dominated by the isolated flat-bands with projected interactions requires a fundamentally new theoretical formulation. The interactions renormalize the bandwidth, the current operator and the diamagnetic susceptibility, and must necessarily contribute to the optical sum-rule. The theoretical formalism to express the low-energy optical spectral weight in terms of the projected many-body degrees of freedom has been developed previously \cite{DMDC1,DMDC2}. However, computing this weight involves knowing the two-particle correlation functions in the non-perturbative many-body ground state (see Eq.~\ref{kxxmom}), which is challenging.

Here we perform an exact analytical computation of the full low-energy optical spectral weight with respect to the ground-state in a strong-coupling limit, without the need for any further approximations. 
One crucial difference between our approach and the traditional route that is well developed for weakly-correlated electrons, is that for the latter it is often sufficient to obtain the interaction-renormalized bands first and compute their optical response (see e.g. \cite{Chaudhary2022shift}), where one expects that the spectral weight is simply related to the renormalized band-parameters. However, in the strong-coupling limit and using our many-body approach, we find that the emergent symmetries of the low-energy spectrum as well as the ground-state projectors have a non-trivial impact on the optical spectral weight.

{\bf Theoretical Framework.-} Our focus in this paper will be on a modified {\it partial} sum-rule that allows us to make {\it exact} statements about the optical spectral weight for moir\'e graphene and beyond. It is possible to obtain an exact relationship between the optical spectral weight associated with the isolated interacting flat bands ($\mathbb{S}$), integrated upto a cutoff energy $E_w$ that lies in the bandgap (Fig.~\ref{fig:schematic} a), and the low-energy diamagnetic response \cite{DMDC1}:
\beq
   {\mathbb{S}} =  \frac{\hbar^2}{e^2}\int_0^{E_w} d\omega ~\tn{Re} [\sigma^{\tn{eff}}_{\tn{xx}}(\omega)] &=& \frac{\pi \hbar^2}{2 e^2 } \langle \kxx \rangle,
\label{eq:partial_f_sum}
\eeq
where $\sigma^{\tn{eff}}_{\tn{xx}}(\omega)$ and $\kxx$ represent the effective longitudinal optical conductivity and diamagnetic response, respectively. It is important to note that both of these quantities are defined in terms of the {\it a priori} unknown ``renormalized" operators. We will focus on density-density interactions, $V$, projected to a set of narrow electronic bands with bandwidth, $W$, that are well isolated by an energy bandgap, $E_{\tn{band}}$, from the remote bands. 

In the limit $W\lesssim V\ll E_{\tn{band}}$, 
the unitary transformation within the low-energy many-body Hilbert space,  $\mathcal{U}_\alpha \equiv e^{i \alpha \mathbb{P}\hat{\mathcal{X}}\mathbb{P}}$, is the effective gauge-transformation tied to the emergent $U(1)$ charge conservation, where $\mathbb{P}=\bigg(\sum_{m\in\tn{active}}|u_m\rangle\langle u_m|\bigg)$ projects to the set of active bands with Bloch functions $|u_m\rangle$, and $\hat{\mathcal{X}}=\sum_\l x^{\phantom\dagger}_\l c^\dagger_\l c^{\phantom\dagger}_\l$ represents the projected many-body position operator; see Appendix for discussion of spectral weight at full band-filling. The effective diamagnetic response (Eq.~\ref{eq:partial_f_sum}) takes the exact form,
\beq
     \langle \kxx \rangle &=& \lim_{\alpha\rightarrow 0} \partial^2_{\alpha}\tn{Tr}\left(\hat\rho~ \mathcal{U}_\alpha~ \mathbb{P} \mathcal{H} \mathbb{P} ~\mathcal{U}_\alpha^\dag \right) \nonumber\\
     &=& -\tn{Tr}\left( \left[\mathbb{P}\hat{\mathcal{X}}\mathbb{P},\left[\mathbb{P}\hat{\mathcal{X}}\mathbb{P},\hat\rho\right]\right]\mathbb{P}\mathcal{H}\mathbb{P}\right),
     \label{kxx_unitary}
\eeq
where $\hat\rho$ is the thermal density matrix. In what follows, we will mostly focus on this response function at zero temperature, where the density matrix reduces to the projector, $\hat\rho_G = |\Psi \rangle \langle \Psi|$, with $|\Psi\rangle$ the many-body ground state wavefunction, but also comment on the results at a finite temperature; see Appendix. The expectation values in Eq.~\ref{kxx_unitary}  
are to be evaluated in the many-body state associated with the non-perturbative Hamiltonian, $\mathcal{H}$, which involves both single and two-particle correlation functions \cite{DMDC1}. Evaluating the latter, without making any further approximations, is {\it a priori} a non-trivial task. Remarkably, we will be able to make a number of {\it exact} statements about $\mathbb{S}$ at zero temperature in the strong-coupling limit of TBG. 

{\bf Twisted bilayer graphene.-} Since the original discovery of superconductivity \cite{Cao2018b} and interaction-induced insulators \cite{Cao2018} in magic-angle twisted bilayer graphene, a plethora of experimental \cite{Yankowitz_2019,Lu2019,Park_2021,Zondiner_2020,uri2020mapping,saito2020independent,saito2021isospin,Cao_2021,liu2021tuning,Das_2021,rozen2021entropic,Serlin900,Sharpe_2019,Stepanov_2020,Wu_2021,Saito_2021Hofstadter,nuckolls2023quantum,Grover2022mosaic,Yu2022hofstadter,Yu2022skyrmion,Morisette2022Hunds,Tseng2022nu2QAH,Choi2019,Oh2021unconventional,Nuckolls2020strongly,Xie2021fractional,Diez-Merida2021diode,Jiang2019charge,Arora2020SC,Kerelsky2019,choi2021correlation,Xie2019stm,Wong_2020,pierce2021unconventional} and theoretical works~\cite{po2018,xie2020weakfield,XieSub,bultinck_ground_2020,liu2021theories,2020CeaGuinea,Zhang2020HF,ochi2018,KangVafekPRL,Kang2020,vafek2020RG,Liu2021nematic,dodaro2018,TBG4,TBG5,TBG6,SoejimaDMRG,kwan_kekule_2021,PotaszMacDonaldED,zhang2021momentum,klebl2021,shavit2021theory,wu2018phonon,lian2019phonon,wu2019phononlinear,lewandowski2021,Bultinck2019mechanism,hejazi2021,parker2020straininduced,thomson2021,Christos_2020,khalaf2021charged,Chatterjee2020skyrmionic,cea2021electrostatic} have attempted to unravel different aspects of the phenomenology.  In this context, if one initially ignores the single-particle dispersion \cite{Bistritzer2011} and in the ``chiral'' limit \cite{tarnopolsky_origin_2019}, the strongly-coupled limit bears resemblance to the classic quantum Hall ferromagnet \cite{bultinck_ground_2020,TBG4} and yields a degenerate manifold of flavor-polarized insulators at commensurate integer fillings (Fig.~\ref{fig:schematic}a, b). Even though there is an interaction-induced insulating gap, $\Delta_{\tn{ins}}$, and no sub-gap optical spectral weight, a finite value for the spectral weight $\mathbb{S}$ signals transitions between the ``filled" and ``empty" states which have a non-zero dipole matrix element (Fig.~\ref{fig:schematic}a). While computing the detailed frequency-dependent optical conductivity reliably is exceedingly difficult, this work establishes a number of exact statements regarding the partial optical spectral weight for these many-body insulating ground states (Fig. \ref{fig:schematic}b). Our first key observation is that due to the emergent symmetry associated with the chiral flat-band limit, $\mathbb{S}=0$. Incorporating perturbations that lift the degeneracy in this limit leading to specific candidate insulating states \cite{bultinck_ground_2020,TBG4,KangVafekPRL,kwan_kekule_2021}, we are able to quantify the features of the resulting many-body states that control the value of $\mathbb{S}$ in realistic parameter regimes of twisted bilayer graphene (Fig.~\ref{fig:schematic}c).

Let us begin with the continuum Hamiltonian for only the isolated bands in TBG directly in momentum-space, $\mathcal{H} = \hk + \hi$, where $\hi = \frac{1}{2A}\sum_{\q} V_\q ~\tilde\rho(\q)\tilde\rho(-\q)$ and $\tilde\rho(\q) = \sum_{\k}\lambda^{\alpha\beta}_\mu(\k,\q) c_{\k,\alpha, \mu}^\dag c^{\phantom\dagger}_{\k - \q, \beta, \mu}$ is the projected density operator. 
Here $\lambda^{\alpha\beta}_\mu(\k,\q) = \langle u_{\k,\alpha,\mu}|u_{\k-\q,\beta,\mu}\rangle$ is the form-factor constructed out of the Bloch functions, and $\mu$ denote the valley/spin quantum numbers while $\alpha, \beta$ denote the sub-lattice indices, and $A$ represents the area. We consider the double-gated Coulomb interaction given by $V_\q= \,V_{0}d\,\tanh\left(qd\right)/q$, with  $V_0=e^2/2\epsilon\varepsilon_0 d$. Taking the dielectric constant to be $\epsilon = 10$ and the screening length $d=25$ nm, we get $V_0 =~18.1$ meV. Clearly, the numerical value of $\mathbb{S}$ will be set by $V_0$, but our interest is primarily in its dependence as a function of other parameters, and on the many-body state itself. We have intentionally left the explicit form of $\hk$ unspecified; a convenient starting point is the Bistritzer-MacDonald (BM) Hamiltonian \cite{Bistritzer2011}, including the various dependencies on twist-angle ($\theta$), ratio of tunneling between AA and AB sites ($w_0/w_1$), and heterostrain ($\varepsilon$). 

We can express $\langle \kxx \rangle$ directly in momentum-space. Focusing on the contribution from $\hi$,
\beq
    \langle \kxx \rangle = \frac{1}{A}\sum_{\q,\k,\k'}  V_\q  \hat{{\cal{D}}}_{x}^2 \left[ \lambda_\mu(\k,\q) \lambda_\nu(\k',-\q)\right]_{\alpha\beta,\gamma\delta}  \times\nn\\
    \langle c_{\k,\alpha,\mu}^\dag c_{\k - \q, \beta,\mu}^{\phantom\dagger}  c_{\k',\gamma,\nu}^\dag c_{\k' + \q, \delta,\nu}^{\phantom\dagger}\rangle,
    \label{kxxmom}
\eeq
where ${\cal{D}}_{x}$ is a covariant derivative acting on both $\k$ and $\k'$ of the form-factors as,
\beq
&&
[\hat{\mathcal{D}}_{x}\lambda_\mu(\k,\q)]_{\alpha\beta} =  (\partial_{k_x} \delta_{\alpha \alpha'} \delta_{\beta \beta'}- i \mathcal{A}_{\k,\alpha\alpha',\mu}^x  + i\mathcal{A}_{\k-\q,\beta\beta',\mu}^x  ) \lambda_\mu^{\alpha' \beta'}(\k,\q), \nn
\eeq
with $\mathcal{A}_{\k,\alpha \alpha',\mu}^\nu = i \langle u_{\k,\alpha,\mu} | \partial_{k_\nu} u_{\k,\alpha',\mu}\rangle$ the multi-orbital Berry connection for valley $\mu$. The bare contribution from $\hk$ follows in a more straightforward fashion \cite{si}. 

{\bf Exact results at integer fillings in the chiral flat-band limit.-} The key advantage of starting from this limit is two-fold: (i) the candidate ground states for the projected Hamiltonian at these fillings are well understood both theoretically \cite{bultinck_ground_2020,TBG4,KangVafekPRL,kwan_kekule_2021} and experimentally \cite{Zondiner_2020,Wong_2020,nuckolls2023quantum}, and (ii) the many-body states themselves are ``Slater-determinant" like, that enable us to compute the above correlation functions in Eq.~\ref{kxxmom} using Wick's theorem. 
At the experimentally relevant integer fillings, we consider a variety of candidate insulating states with the following single-particle density matrix, $\langle c_{\k}^\dag c^{\phantom\dagger}_{\k'} \rangle \equiv {\cal{P}}_{\k,\k'}$,
where $c_\k$ is defined now in the {\it sub-lattice basis} and we have suppressed the explicit dependence on the other quantum numbers. We have analyzed a variety of integer fillings in this work \cite{si}, but for the sake of brevity focus only on $\nu=-2$ here.  
In Table \ref{tab:order_para} in the appendix, we list the ${\cal{P}}_{\k,\k'}$'s for a few representative states along with their dependence on the different quantum numbers at $\nu=-2$: valley polarized (VP), valley Hall (VH), quantum anomalous Hall (QAH), intervalley coherent states (KIVC and TIVC), and incommensurate Kekul\'e spiral (IKS). In the chiral-flat band limit, the Hamiltonian has an exact $U(4)\times U(4)$ symmetry and all of the above states (except IKS) are degenerate ground states. Small deviations from the chiral-flat limit break this degeneracy in a way that favors intervalley coherent states. However, with the inclusion of heterostrain, the bandwidth is dramatically enhanced and translation symmetry breaking orders like IKS are found to be favorable from Hartree-Fock mean-field and DMRG studies \cite{kwan_kekule_2021,wang2022kekule,wagnerGlobal2022}.

In order to calculate $\langle \kxx \rangle$, we first discuss the form of the projected position operator $\mathbb{P}\hat{\mathcal{X}}\mathbb{P}$ in TBG. In the sub-lattice basis of TBG, the projected position operator can be written as,
$\mathbb{P}\hat{\mathcal{X}}\mathbb{P} = \sum_{\k} c^\dag_{\k,\alpha,\mu}  (i \delta_{\alpha,\alpha'}  \partial_{\k_x}  + \mathcal{A}_{\k, \alpha \alpha',\mu}) c^{\phantom\dagger}_{\k, \alpha',\mu}$.
We adopt a previously discussed gauge-fixing scheme \cite{bultinck_ground_2020} whereby there is no non-abelian Berry connection between the two sub-lattices. We therefore define $\mathcal{A}_{\k, \alpha \alpha',\mu} = \delta_{\alpha,\alpha'} \mathcal{A}_{\k, \alpha,\mu}$. For the two sublattices of the same valley, we have $\mathcal{A}_{\k, 1,\mu} = -\mathcal{A}_{\k, 2,\mu}$. For the two opposite valleys, from the combination of $C_{2z}$ and particle-hole symmetry, we have $\mathcal{A}_{\k, 1,1} = \mathcal{A}_{\k, 2,2}$ and $\mathcal{A}_{\k, 2,1} = \mathcal{A}_{\k, 1,2}$. We further define $\mathcal{A}_{\k, 1,1} = \mathcal{A}_{\k}$ such that,
\beq
\mathbb{P}\hat{\mathcal{X}}\mathbb{P} = \sum_{\k} \hat{c}_{\k}^\dag (i \partial_{\k_x} \mathbf{1} + \mathcal{A}_{\k}^x \sigma_z\tau_z)\hat{c}^{\phantom\dagger}_{\k},
\eeq
where $\sigma$ and $\tau$ act on the sub-lattice and valley basis, respectively.
\begin{figure*}[pth!]
\centering
\includegraphics[width=1.0\linewidth]{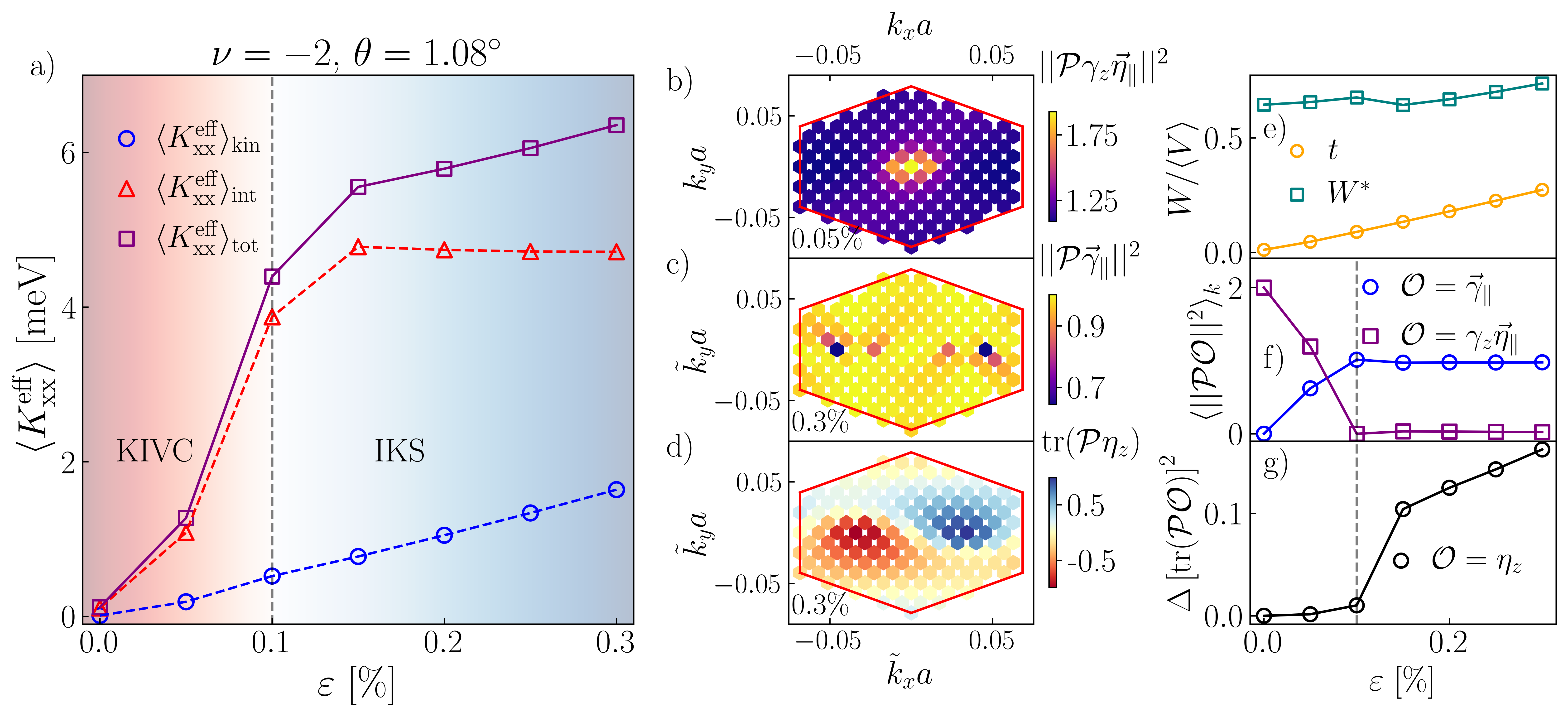}
\caption{ (a) The effective diamagnetic response (see Eq.~\ref{eq:partial_f_sum}) evaluated at $T=0$ and for $\nu = -2$ as a function of strain within a fully self-consistent Hartree-Fock computation. The dashed line demarcates the transition from a KIVC to an IKS insulator. (b)-(d) Order parameters characterizing the ground state for $\varepsilon=0.05\%$ and $\varepsilon=0.3\%$, respectively. In (c) and (d), we use the momentum-boosted mini-Brillouin zone. To account for the gauge dependence of $\vec{\gamma}_{\parallel}=(\gamma_x,\gamma_y)$ and the phase of the IVC order parameter, we plot $||\mathcal{PO} ||^2 =|\mathrm{tr}_{\gamma \eta}(\mathcal{PO})^2|$ in (b), (c), where $\mathrm{tr}_{\gamma \eta}$ is the partial trace over valley and Chern pseudospin and $|\cdot|$ is the Frobenius norm.  
(e) Evolution of $t/\langle V\rangle$ and $W^*/\langle V\rangle$, where $W^*$ represents the interaction induced bandwidth with increasing strain \cite{si}. (f) The momentum-averaged order parameters in (b) and (c) with increasing $\varepsilon$. 
(g) Variation of the momentum-averaged fluctuation in the valley polarization as a function of strain. 
}
\label{fig:approx_kxx}
\end{figure*}
Consider the projector at $\nu = -2$, $\rho_G$, for the degenerate ground states associated with the chiral flat-band limit, 
\begin{subequations}
\beq
\hat{\rho}_G &\equiv&
\bigotimes_{\k} c_{\k,\alpha}^\dag c_{\k,\beta}^\dag  |0\rangle_{\k} \left[\hat{\mathcal{T}}_{\k}\right]^{\alpha\beta}_{\gamma\delta}{}_{\k}\langle 0|  c^{\phantom\dagger}_{\k,\gamma} c^{\phantom\dagger}_{\k,\delta},\\
\left[\hat{\mathcal{T}}_{\k}\right]^{\alpha\beta}_{ \gamma\delta} &=& \frac14 \delta_{\k,\k'} \varepsilon_{\sigma \delta \gamma} \varepsilon_{\sigma \delta' \gamma'}  \left[\mathcal{P}_{\k,\k'}^T\right]_{\alpha\delta'} \left[\mathcal{P}_{\k,\k'}^T\right]_{\beta\gamma'},
\eeq
\end{subequations}
where $|0\rangle_{\k}$ denotes the empty state at momentum $\k$, and $\alpha,\beta,\gamma,\delta$ denote the valley-sublattice-spin indices that are being summed over, with the spin polarization fixed by the structure of the projectors. The matrix $\left[\hat{\mathcal{T}}_{\k}\right]_{\alpha\beta,\gamma\delta}$ is constructed out of the single-particle density matrices, $\mathcal{P}$. 

A key observation is that $\mathcal{P}_{\k,\k'}$ commutes with $\sigma_z\tau_z$, and hence $\left[\mathbb{P}\hat{\mathcal{X}}\mathbb{P}, \hat\rho_G \right] = 0$, implying $\mathbb{S}\propto\langle \kxx \rangle = 0$. From the point of view of the representation of the many-body states in Fig. \ref{fig:schematic}b, a vanishing $\mathbb{S}$ follows from the fact that the action of $\mathcal{U}_\alpha$ leaves the Bloch vectors invariant. 
Thus, all of the degenerate correlated many-body states associated with the chiral flat-band limit have a vanishing $\mathbb{S}$ at $T=0$. This implies that there are no allowed optical transitions --- associated with the usual spin and valley diagonal current operators ---between the filled and empty states. In other words, for these many-body states, $\tn{Re}[\sigma^{\tn{eff}}_{\tn{xx}}(\omega)]=0$ for all $\omega<E_w$ (i.e. as long as inter-band transition to the dispersive bands is forbidden). Notably, there are other examples where $\mathbb{S}=0$, including for instance, in the theory for the {\it projected} lowest (spinless) Landau-level \cite{DMDC2}. In this example, the $f-$sum-rule is saturated at the cyclotron resonance as a result of Kohn's theorem \cite{kohn1961,yip89}, which lies outside of the low-energy manifold. In the projected lowest Landau-level theory, the emergent dipole-conservation forbids any ``intra-band" optical transitions. 
In the limit of full polarization associated with the generalized ferromagnetic states that appear at various integer fillings in TBG, any optical transition requires flipping a quantum number that is conserved by the current operator, forbidding a dipole matrix element.

{\bf Approximate results at integer fillings away from the chiral flat-band limit.-} 
To quantify how experimentally relevant deviations from the exactly solvable chiral-flat band limit enhance $\mathbb{S}$, we evaluated $\langle \kxx \rangle$ approximately (Fig.~\ref{fig:approx_kxx}a)) using the ground states of $\mathcal{H}$ obtained self-consistently from Hartree-Fock as a function of strain $\varepsilon$  (see \cite{si} and \cite{kwan_kekule_2021,nuckolls2023quantum}). Our results converge with both increasing system size and number of Umklapp processes in the moir\'e Brillouin zone. 
In this procedure, the set of active flat bands are renormalized by interactions with states from the remote bands giving rise to an enhanced bandwidth $W^*$ \cite{si}. We keep track of all flavors and consider all possible translation symmetry breaking order parameters in the intervalley sector, $\langle c^\dagger_{\vec{k} -\tau \vec{q}/2,\tau,\sigma,s} c_{\vec{k}-\tau'\vec{q}/2,\tau', \sigma',s'} \rangle = \mathcal{P}_{\tau,\sigma,s,\tau',\sigma',s'}(\vec{k})$. At low $\varepsilon$, the state at $\nu=-2$ does not break translation symmetry and closely resembles $\mathcal{P}_{\rm KIVC}(\vec{k})$ but acquires a weak momentum dependence over the Brillouin zone (Fig.~\ref{fig:approx_kxx}b)). Further increasing $\varepsilon$ enhances $H_{\mathrm{kin}}$, which includes terms that anticommute with the KIVC order parameter leading to a suppression of IVC order in favor of states that have coherence between the different Chern sectors \cite{bultinck_ground_2020}. At $\nu=-2$, the Hartree renormalization of the dispersion additionally introduces momentum-dependent features which favor translation symmetry breaking at an incommensurate wavevector $\vec{q}_{\mathrm{IKS}}$ in the intervalley channel \cite{kwan_kekule_2021, wagner_global_2021,nuckolls2023quantum}. The resulting state consists of an equal superposition of Chern-zero and Chern-coherent components with IVC order that preserves time-reversal symmetry. In this state, there is a modified translation symmetry that allows for a redefinition of the Brillouin zone in terms of the valley boosted momentum $\vec{\tilde{k}} = \vec{k} - \tau_z \vec{q}_{\mathrm{IKS}} /2$. We show the Chern coherent component, and the momentum dependent valley-polarization in Fig.~\ref{fig:approx_kxx}c-d) respectively. While the Chern-coherent component has a non-zero average across momentum space, the valley polarization averages to zero with strong momentum dependent features that correspond to the minima of the renormalized dispersion. Both of these features control the deviations away from $\mathbb{S}=0$.

The density matrix of the IKS states can be written as,
\beq
\rho_G \equiv
\bigotimes_{\tilde{\k}} c_{\tilde{\k},\alpha, \uparrow}^\dag c_{\tilde{\k},\alpha', \downarrow}^\dag |0\rangle_{\tilde{\k}} \left[\hat{\mathcal{P}}_{\tilde{\k}}\right]_{\alpha\beta}\times \left[\hat{\mathcal{P}}_{\tilde{\k}}\right]_{\alpha'\beta'}{}_{\tilde{\k}}\langle 0| c_{\tilde{\k},\beta', \downarrow} c_{\tilde{\k},\beta, \uparrow}
\eeq
where the momentum $\k = \tilde{\k} + \q \tau_z/2$, and $\uparrow, \downarrow$ denote the spin; summation over valley-sublattice indices $\alpha, \beta$ is implicit. $\hat{\mathcal{P}}_{\tilde{\k}}$ is related to the single-particle density-matrix, $\hat{\mathcal{P}}_{\tilde{\k}} = \mathcal{P}_{\k,\k'}^T \delta_{\k,\tilde{\k}+ \q \tau_z/2} \delta_{\k',\tilde{\k}+ \q \tau_z/2}$. This ansatz would correspond to the Bloch vector lying in the XY-plane. There are two contributions to $\left[\mathbb{P}\hat{\mathcal{X}}\mathbb{P},\rho_G \right]$, that arise from (i) the momentum dependence of $n_{\tilde{\k}}$ and $m_{\tilde{\k}}$ (see Table \ref{tab:order_para} in Appendix), and (ii) the commutator of $\gamma_{x,y}$ with $\sigma_z\tau_z$, respectively. More generally, as long as the Bloch vector in Fig.~\ref{fig:schematic}c is not aligned towards the pole (as in the previous  strong-coupling limit), or develops a non-trivial momentum dependence, the optical spectral weight $\mathbb{S}\neq0$. It is readily seen that $\left[\mathbb{P}\hat{\mathcal{X}}\mathbb{P},\rho_G \right] \neq 0$ for these IKS states. 

 We can track the evolution of the many-body state with increasing $\varepsilon$ by examining the momentum dependent features of the projection of $\mathcal{P}$ along different operators, $\mathcal{O}$, that act on valley, spin and sublattice, respectively. To quantify the momentum dependencies of the order parameters, we plot them in the Brillouin zone in Fig.~\ref{fig:approx_kxx}b,c and their average, $\langle||\mathcal{PO}||^2\rangle_k$, in Fig.~\ref{fig:approx_kxx}f. Note we avoid both gauge ambiguities and ambiguities associated with the U(2) spin rotations of the IVC order parameter by calculating the Frobenius norm of the partial trace of the state over Chern and valley degrees of freedom, $||\cdot||^2$, in b) and c). At small $\varepsilon$, the state $\mathcal{P}$ develops contribution parallel to the semimetallic order parameter $\mathcal{O}_{SM} \propto \gamma_x,\gamma_y \propto t/ \langle V \rangle$, where $\langle V \rangle = \frac{1}{2A}\sum_{\vec{q},\vec{k}} V_{\vec{q}} \mathrm{tr}(\lambda(\vec{k},\vec{q})\lambda(\vec{k},-\vec{q}))$ measures the strength of the interaction \cite{bultinck_ground_2020}. Eventually, the KIVC order is subsumed by IKS and this average saturates when $\mathcal{P}$ lies fully in the XY plane of the Chern Bloch sphere in Fig.~\ref{fig:schematic}c.  The development of this component of $\mathcal{P}$ coincides with the enhancement of $\langle \kxx \rangle$, in accordance with the contribution highlighted in (i) above. We analyze the momentum variation of the projection of $\mathcal{P}$ onto $\mathcal{O}$, which we probe via $\Delta \left[\mathrm{tr}(\mathcal{PO})\right]^2 = \sum_{\vec{k}}[\mathrm{tr}(\mathcal{PO}) - \sum_{\vec{k}}(\mathrm{tr}\mathcal{PO})]^2$, for the valley polarization, $\eta_z$, in Fig.~\ref{fig:approx_kxx}g. Even though the order parameters in Fig.~\ref{fig:approx_kxx}f saturate as a function of strain, the fluctuations tied to the valley polarization in Fig.~\ref{fig:approx_kxx}g increase monotonically, which is responsible for the continued increase of $\kxx$ in Fig.~\ref{fig:approx_kxx}a \cite{si}.

{\bf Outlook.-} Our approach enables a systematic understanding of the factors associated with the low-energy theory that contribute to the finite optical spectral weight in interacting topological flat-band systems. Starting from the chiral flat-band limit of TBG at commensurate integer fillings, where the low-energy spectral weight vanishes, we compute the effects of realistic strain, twist-angle, and most importantly the many-body ground-state on the optical spectral weight. Even in the strong-coupling limit, doping away from the insulating (integer) fillings by an amount $\delta\nu$ enhances $\delta\mathbb{S}\propto\delta\nu$. Interestingly, the IKS state, which has been observed experimentally at $\nu=-2$ \cite{nuckolls2023quantum}, is particularly effective in enhancing this spectral weight compared to other intervalley coherent states (e.g. KIVC). A larger available optical spectral weight does not guarantee higher $T_c$ when doped away from the insulating limit, but it is beneficial since only a small fraction typically condenses into the superfluid stiffness. 

There are several interesting open questions. In light of the recent progress with the ``heavy-fermion'' perspective on this problem \cite{Song2022heavy} and experiments on twisted trilayer graphene \cite{Park2021TTGSC,Hao2021TTGelectric}, it will be interesting to extend the theoretical formalism to include the contributions to the optical spectral weight from some of the closest (dispersive) remote bands. In MATBG, the closest remote band has a band gap $E_{\tn{gap}}\sim O(50-100~\tn{ meV})$, and we expect its contribution to the optical spectral weight to be suppressed in $V/E_{\tn{gap}}$. For twisted trilayer graphene, the additional dispersive Dirac-like bands at low-energies contribute to the optical spectral weight, which necessitates the development of methods to include their contribution in an energy-scale dependent fashion. Moreover, computing the contribution to the optical spectral weight due the electron-phonon interactions, where the phonons have their own non-trivial dynamics, also remains an important future direction to explore. Finally, while the exact nature of the metallic normal state obtained upon doping the correlated insulators in these moir\'e systems remains unclear, developing a systematic approach to compute the Drude weight is an exceptionally important exercise. Given the intertwined nature of the robust superconducting phases and IKS order in the vicinity of $\nu=-2$, it is natural to ask whether the reduced low-energy optical spectral weight discussed in this paper and the extended regime of $T-$linear resistivity observed experimentally down to low temperatures \cite{TBGSM,Jaoui2022}, fundamentally limits how large $T_c$ can be in twisted bilayer graphene \cite{PALNR,SSAJM}. 

Recent experimental measurements of the superfluid stiffness in moir\'e graphene \cite{tanaka2024kinetic,banerjee2024superfluid} fall well within the upper bound suggested from our computations of the many-body optical spectral weight. Building on recent technical advances geared towards moir\'e systems \cite{banerjee2024superfluid,basov,kipp2024cavity,koppens}, future experiments will likely enable measurements of the low-frequency optical conductivity in these heterostructures in low and high-strained samples, which will help make detailed connections to the theoretical results contained here.
Beyond moir\'e graphene, superconductivity has been recently reported in twisted WSe$_2$ \cite{xia2024unconventional,guo2024superconductivity}. Given the close connection between a correlated insulator at integer filling and the superconductor \cite{xia2024unconventional}, a natural extension of the current setup might help reveal the factors that fundamentally limit $T_c$ in this platform.

{\it Acknowledgements.-} We thank E. Berg, B.A. Bernevig, Y.H. Kwan and M. Randeria for a number of useful  discussions. 
JFMV and DC are supported in part by a NSF CAREER grant (DMR-2237522) and a Sloan Research Fellowship to DC. DM is supported by a Bethe postdoctoral fellowship at Cornell University.

\bibliography{refs.bib}

\clearpage
\begin{appendix}

\begin{widetext}
\begin{center}
  \textbf{\large End Matter}
\end{center}
\end{widetext}

{\bf Optical spectral weight in fully-filled bands.-} At full filling of the bands in the projected limit (e.g. at $\nu=4$ in MATBG) with a band-insulating ground-state, $\langle\kxx\rangle=0$ vanishes trivially. This follows from the observation that $\mathcal{U}_\alpha = e^{i \alpha \mathbb{P}\hat{\mathcal{X}}\mathbb{P}}$ acting on the fully filled states only gives rise to a phase factor and there is no ``intra-band" transition within the low energy manifold. In other words, for the full filling, $\left[ \mathcal{U}_\alpha, \hat\rho_G \right] = 0$, and $\langle \kxx \rangle$ vanishes according to Eq.~\ref{kxx_unitary}. 

{\bf Optical spectral weight at finite temperature.-} At a finite temperature, there are two interesting limits we can consider: (i) ``high" temperature, where $V \ll T \ll E_{\text{band}}$, and (ii) a low temperature, where $T \ll V $. At the leading order in a high-temperature expansion ($\beta\rightarrow0$), the thermal density matrix in the low-energy subspace, $\hat{\rho} \sim \mathbf{1}$ leading to $\left[\mathbb{P}\hat{\mathcal{X}}\mathbb{P},\hat{\rho} \right] = 0$, and a vanishing $\mathbb{S}$. This arises due to the vanishing available phase-space at low-energy for transitions due to the equal-weight thermal population over all the many-body states. 

\begin{table}[!ht]
    \centering
    \setlength\extrarowheight{2mm}
    \begin{tabular}{|c|c|}
   \hline
       {\bf Correlated }  &  {\bf Single-Particle}\\
      {\bf States} & {\bf Density-matrix} (${\cal{P}}_{\k,\k'}$)\\
       \hline \hline
        {\bf VP} & $\delta_{\k,\k'}\frac12 \left( \mathbf{1} \pm \tau_z \right) \otimes {\cal{P}}_s$\\
        {\bf VH} & $\delta_{\k,\k'} \frac12 \left( \mathbf{1} \pm \sigma_z\right) \otimes {\cal{P}}_s $\\
        {\bf QAH} & $\delta_{\k,\k'} \frac12 \left( \mathbf{1} \pm \sigma_z \tau_z \right)\otimes {\cal{P}}_s$\\
        {\bf KIVC} & $\delta_{\k,\k'}\frac12  \left[\mathbf{1} + \sigma_y( \cos\phi \tau_x + \sin\phi \tau_y)\right]\otimes {\cal{P}}_s$\\
        {\bf TIVC} & $\delta_{\k,\k'}\frac12  \left[\mathbf{1} + \sigma_x( \cos\phi \tau_x + \sin\phi \tau_y)\right]\otimes {\cal{P}}_s$\\
        \hline \hline
     {\bf IKS} & $\delta_{\k,\tilde{\k}+ \q \tau_z/2} \delta_{\k',\tilde{\k}+ \q \tau_z/2}\frac14  \left(\mathbf{1} + n_{\tilde{\k}} \cdot \gamma\right) \left(\mathbf{1} + m_{\tilde{\k}} \cdot \eta\right)$\\
             \hline
    \end{tabular}
    \caption{The single-particle density matrix, ${\cal{P}}_{\k,\k'}$, for different Slater-determinant-like states at $\nu=-2$. The matrices $\tau$ and $\sigma$ act on the valley and sub-lattice basis, respectively. Here $\gamma = (\sigma_x,\sigma_y\tau_z, \sigma_z \tau_z)$, $\eta = (\sigma_x\tau_x, \sigma_x \tau_y, \tau_z)$ and $n_{\tilde{\k}}$ lies in the XY plane. $\mathcal{P}_s$ denotes the density-matrix in spin space, 
    $\mathcal{P}_s =\frac12( \mathbf{1} + s_{\vec{n}_s})$, $\vec{n}_s$ being the polarization of the spin. The states included in the upper block are a subset of the degenerate ground states for the projected interaction-only model in the chiral flat-band limit. The IKS state in the lower block is the Hartree-Fock ground state in the presence of a single-electron bandwidth, including effects of strain. See \cite{si} for a discussion of correlated states at other fillings. 
    }
    \label{tab:order_para}
\end{table}
In the low temperature regime, we expect that the correction to the spectral weight is $O(e^{- \Delta_{\text{ins}}/T})$, governed by the excited state above the many-body gap $\Delta_{\text{ins}} \sim O(V)$. We note here an important difference in the low-temperature behavior of the spectral weight between the chiral-flat band of MATBG and the lowest Landau-level. For the latter, the dipole conservation is a symmetry of the low-energy Hamiltonian and therefore $\mathbb{S} \sim 0$ for $T \ll \hbar \omega_c$, and $\mathbb{S} \sim O(e^{- \hbar \omega_c/T})$ governed by cyclotron frequency. In the chiral-flat band limit of MATBG, the dipole conservation is only a property of the ground state manifold and not the low-energy Hamiltonian, so the thermal density matrix at a finite but low temperature does not have a vanishing $\mathbb{S}$.

{\bf Perturbative corrections to optical spectral weight away from chiral limit.-}
Consider a perturbation in the form of a single-particle kinetic energy, $\hk~(\sim O(t))$. When the characteristic scale of the bare electron kinetic energy, $t\ll\Delta_{\rm{ins}}$, the correlated insulator is expected to remain stable. However, the ground state wavefunction does not necessarily remain unperturbed. Even for an infinitesimal $t$, the states acquire a correction that is $O(t)$. However, as we now argue, the correction to $\mathbb{S}$ is only $O(t^2/V)$. Recall that the quantity of interest is $\langle \psi_{\tn{new}} |[\mathbb{P}\hat{\mathcal{X}}\mathbb{P}, [\mathbb{P}\hat{\mathcal{X}}\mathbb{P},\hi+\hk]] |\psi_{\tn{new}}\rangle $, where $|\psi_{\tn{new}}\rangle$  is the ground state of the perturbed Hamiltonian $\hi+\hk$. The $O(t)$ correction to $\mathbb{S}$ is given by, 
\beq
\delta\mathbb{S} &=& \langle \psi_{0} |[\mathbb{P}\hat{\mathcal{X}}\mathbb{P}, [\mathbb{P}\hat{\mathcal{X}}\mathbb{P},\hi ]] |\psi_{\tn{new}}\rangle 
+ \langle \psi_{\tn{new}} |[\mathbb{P}\hat{\mathcal{X}}\mathbb{P}, [\mathbb{P}\hat{\mathcal{X}}\mathbb{P},\hi ]] |\psi_0\rangle  \nn\\
&+& \langle \psi_{0} |[\mathbb{P}\hat{\mathcal{X}}\mathbb{P}, [\mathbb{P}\hat{\mathcal{X}}\mathbb{P},\hk ]] |\psi_0\rangle,
\eeq
where $|\psi_0\rangle$ is the ground state of the unperturbed Hamiltonian, $\hi$. For the generalized FM states, $|\psi_0\rangle$, in the chiral flat-band limit, we have $[\mathbb{P}\hat{\mathcal{X}}\mathbb{P}, [\mathbb{P}\hat{\mathcal{X}}\mathbb{P},\hi]] |\psi_0\rangle = 0$ and $[\mathbb{P}\hat{\mathcal{X}}\mathbb{P}, [\mathbb{P}\hat{\mathcal{X}}\mathbb{P},|\psi_0\rangle \langle \psi_0|]] = 0$. This immediately leads to the $O(t)$ contribution to $\delta\mathbb{S}$ to vanish. 
Importantly, this suppression is intimately tied to the generalized FM being the un-perturbed ground state. In more general settings, one would otherwise expect $\mathbb{S}\sim O(t)$ from the single-particle bandwidth.

\vspace{20pt} %
\vspace{20pt} %
\vspace{20pt} %
\vspace{20pt} %
\end{appendix}

\clearpage
\renewcommand{\thefigure}{S\arabic{figure}}
\renewcommand{\figurename}{Supplemental Figure}
\setcounter{figure}{0}
\appendix
\pagenumbering{arabic}
\begin{widetext}

\begin{center}
  \textbf{\large SUPPLEMENTARY INFORMATION}\\[.2cm]
  \textbf{\large Low-energy optical sum-rule in moir\'e graphene}\\[.2cm]
  J.F. Mendez-Valderrama, Dan Mao, and Debanjan Chowdhury
  
  {\itshape
        \mbox{Department of Physics, Cornell University, Ithaca, New York 14853, USA.}\\
	
 }
\end{center}

\section{Strained Continuum Model} \label{appendix_model}

In our calculations, the starting point is the Bistritzer-MacDonald (BM) model for twisted bilayer graphene \cite{Bistritzer2011} including strain and interlayer tunneling. We use the following convention for the reciprocal lattice vectors of the underlying graphene lattices:
\begin{equation}
    \vec{G}_1 = \frac{4\pi}{\sqrt{3} a_{\mathrm{Gr}}}\bigg(0 ,-1 \bigg),\,\,\, \vec{G}_2 = \frac{4\pi}{\sqrt{3}a_{\mathrm{Gr}}}\bigg(\frac{\sqrt{3}}{2} ,-\frac{1}{2}\bigg),
\end{equation}
where $a_{Gr}=0.246$nm is the lattice constant of graphene. We can capture the effect of both uniaxial strain on layer $\ell$, $\varepsilon_\ell$, and a rotation angle $\theta_\ell$ by implementing a linear transformation in the small deformation limit \cite{Bi2019} :
\begin{align}
\vec{G}\rightarrow \vec{G}^{\prime} &= M(\theta_\ell, \varepsilon_\ell, \varphi_s ) \vec{G} \\
&= \bigg( R(\theta_\ell) + S(\varepsilon_\ell, \varphi_{s} )\bigg)^{-1} \vec{G},
\end{align}
where $R$ is a rotation matrix and $S$ is the transformation:
\begin{align}
     S(\varepsilon_\ell, \varphi_{s} ) & = \left(\begin{array}{cc}
\varepsilon_{\ell,xx} & \varepsilon_{\ell,xy}\\
\varepsilon_{\ell,xy} & \varepsilon_{\ell,yy}
\end{array}\right)\\
     & = \varepsilon_\ell\left(\begin{array}{cc}
\nu_p\sin^{2}\varphi_{s}-\cos^{2}\varphi_{s} & \left(1+\nu_p\right)\sin\varphi_{s}\cos\varphi_{s}\\
\left(1+\nu_p\right)\sin\varphi_{s}\cos\varphi_{s} & \nu_p\cos^{2}\varphi_{s}-\sin^{2}\varphi_{s}
\end{array}\right).
\end{align}

Here, $\nu_p=0.16$ is the Poisson ratio of graphene. We set the strain direction $\varphi_s = 0$ for the results in the main text. With these conventions we calculate the moir\'e reciprocal lattice vectors by performing a symmetric twist of the graphene layers. The rotation angle for each graphene layer is thus $\theta_1 = -\theta_2 =\theta/2$ with $\theta$ being the twist angle. Similarly, we deal exclusively with uni-axial hetero-strain, so that the strain for each graphene layer, $\varepsilon_\ell$, is related to the relative strain $\varepsilon$ via $\varepsilon = -\varepsilon_2 =\varepsilon/2$. With these parameters, the moir\'e reciprocal lattice vectors are obtained via 
\begin{equation}
    \vec{G}^M_{1,2} = \left( M(\theta_1, \varepsilon_1, \varphi_s ) - M(\theta_2, \varepsilon_2, \varphi_s ) \right) \vec{G}_{1,2}.
\end{equation}
The new moir\'e Brillouin zone (MBZ) in general consists of a deformed hexagon since a finite $\varepsilon$ breaks $C_{3z}$ and $C_{2x}$ . Nevertheless, we can plot quantities in an undeformed Brillouin zone by mapping the $k-$points to the deformed MBZ via the transformation  
\begin{equation*}
     L(\theta, \varepsilon, \varphi_s)=\left( M(\theta_1, \varepsilon_1, \varphi_s ) - M(\theta_2, \varepsilon_2, \varphi_s ) \right) \left( M(\theta_1, 0, \varphi_s ) - M(\theta_2, 0, \varphi_s ) \right)^{-1}.
\end{equation*}

With these conventions, the  strained BM model  takes the form, 
\begin{equation}
    H_{BM}  = H_{\mathrm{Dirac} } + H_{\mathrm{interlayer}},
\end{equation}
where the Dirac part of the Hamiltonian descends from the Hamiltonian of the decoupled graphene sheets and the interlayer contribution captures the leading order contributions to the interlayer hopping processes which connect the $K$ ($K'$) points of the underlying graphene sheets.  Due to strain, the Dirac points are unpinned from the $K$ ($K'$) points as $C_{3z}$ is broken, and they also can shift away from zero energy since $C_{2x}$ is broken; however, the crossing remains protected since $C_{2z}\mathcal{T}$ remains a good symmetry. This changes are captured by the fact that, within the tight binding approximation, the hopping amplitudes change in the presence of strain. At low energies, these changes are reflected in the emergence of a pseudomagnetic field for the Dirac hamiltonian at momentum $k$ for a single layer:
\begin{equation}
   \langle \vec{k},\ell | H_{\mathrm{Dirac} } |\vec{k}'\ell ' \rangle = -\delta_{\vec{k},\vec{k}'}\delta_{\ell,\ell'}\hbar v_f(\tau_z \sigma_x, \sigma_y) \cdot M(\theta_\ell, \varepsilon_\ell, \varphi_s ) \left( \vec{k} - \tau_z \vec{A}_{\ell} - \tau_z M(\theta_\ell, \varepsilon_\ell, \varphi_s )^{-1} \vec{K} \right),
\end{equation}
where $\vec{k}$ is measured w.r.t. the $\Gamma$  point of the original graphene BZ, $\vec{K}=\frac{2\vec{G}_2 - \vec{G}_1}{3}$, and 
\begin{equation}
    \vec{A}_\ell = \beta_A \left( \varepsilon_{\ell,xx} - \varepsilon_{\ell,yy}, -2\varepsilon_{\ell,xy} \right),
\end{equation}
with $\beta_A = 3.14 \frac{\sqrt{3}}{ 2 a_{\mathrm{Gr}} } $. Furthermore, the  Dirac velocity is such that $\hbar v_f / a_{\mathrm{Gr}} = 2.4$24eV. Additionally, the interlayer hopping is given by:
\begin{equation}
 \langle \vec{k}, 1 |H_{\mathrm{interlayer}} |\vec{k}', 2'\rangle= w\delta_{\vec{k}-\vec{k}',\vec{0}} T_1 + w\delta_{\vec{k}-\vec{k}',\vec{G}^M_1} T_2 + w \delta_{\vec{k}-\vec{k}',\vec{G}^M_2} T_3,
\end{equation}
where the $T$  matrices act on the sublattice degrees of freedom and correspond to 
\begin{align}
T_{1} & = \kappa \sigma_{0}+\sigma_{x}\\
T_{2} & = \kappa \sigma_{0}+\cos\left(\frac{2\pi}{3}\right)\sigma_{x}+\sin\left(\frac{2\pi}{3}\right)\sigma_{y}\\
T_{3} & = \kappa \sigma_{0}+\cos\left(\frac{2\pi}{3}\right)\sigma_{x}-\sin\left(\frac{2\pi}{3}\right)\sigma_{y},
\end{align}
where we take $w=110$meV and  $\kappa = 0.75$  unless otherwise stated. This model and parameters are the basis of all the calculations shown in the main text. 

\section{Gauge Fixing} \label{appendix_gauge_fixing}

We fix the phase of the wavefunctions before carrying out the Hartree-Fock procedure outlined in the following section. This also expedites the calculation of the wavefunctions themselves as we can use symmetries of the Hamiltonian to reconstruct the wavefunctions of one valley from the opposite valley. For convenience, we pick a gauge where the wavefunctions are periodic in momentum space and locally smooth. Throughout the main text, we work in the sublattice basis. To fix the gauge we use an emergent anti-unitary particle-hole symmetry that is present in the BM model at small angles. As such, we first assume the small angle approximation for the BM model. This approximation amounts to neglecting the twist angle dependence of $H_{\mathrm{Dirac}}$. The procedure that we use for gauge fixing is then outlined as follows:
\begin{enumerate}
    \item Fix the representation of $C_{2z}\mathcal{T}$ in the band basis to fix the phases of the wavefunctions for valley $\tau=+$ up to a sign.
    \item Fix the additional sign ambiguity by imposing the sublattice chiral symmetry $\mathcal{C}_{\mathrm{sub}}$ in the band basis which eliminates the sign ambiguity between the components of the wavefunctions in different bands. Following this step we perform a basis transformation to the sublattice basis.
    \item Obtain the wavefunction of valley $-$ by fixing the representation of the emergent particle-hole symmetry in the small angle approximation, $\mathcal{P}\mathcal{T}$, in the sublattice basis.
    \item Make the gauge locally smooth by calculating the wavefunction overlaps $\lambda_\tau^{\sigma,\sigma'}(\vec{k},\vec{k}')$ for nearby $\vec{k}-$points and locally imposing that $\lim_{\vec{q} \rightarrow 0} (\lambda_\tau^{+,+}(\vec{k},\vec{q})-\overline{\lambda_\tau^{-,-}(\vec{k},\vec{q})})= 0$, with the form factors $\lambda_\tau^{\sigma,\sigma'}(\vec{k},\vec{k}')$ defined in the main text.
\end{enumerate}

Conditions 1--3 fix the relative phases of wavefunctions in both sublattices and both valleys at each momentum $\vec{k}$ as well as form momentum $-\vec{k}$ by time-reversal symmetry. Condition 4 locally matches between different $\vec{k}$ points to make the gauge locally smooth. Due to the topology of the bands, the gauge cannot be chosen to be globally smooth and in general singularities will appear at degenerate points. We avoid these points in our $\vec{k}-$point sampling by shifting the momentum grid by $10^{-7}/a_M$ in our calculations.

To carry out the gauge-fixing procedure outlined above we choose the following representations for the symmetries for the first-quantized Hamiltonian in the band basis
\begin{equation}
    B^{C_{2z}\mathcal{T} } (\vec{k}) = n_z \mathcal{K}, \,\,\,\, B^{ \mathcal{C}_{\mathrm{sub}}} (\vec{k}) = n_x,
\end{equation}
where $\mathcal{K}$ is complex conjugation and $n_{(x,y,z)}$ are the Pauli matrices in the band basis and $\mathcal{C}_{\mathrm{sub}}$ is a symmetry only when $\kappa=0$. Having fixed the gauge within a single valley, we transform to the sublattice basis by applying the unitary matrix
\begin{equation}
    U_{\sigma,n}(\vec{k}) = \frac{1}{\sqrt{2}}\left[\begin{array}{cc}
1 & 1\\
1 & -1
\end{array}\right]_{\sigma,n}.
\end{equation}
In the sublattice basis, we then choose a representation for the emergent particle-hole symmetry 
\begin{equation}
    B^{\mathcal{P}\mathcal{T} } (\vec{k}) =  -i \tau_y \sigma_y .
\end{equation}
This procedure completely fixes the gauge of the wavefunctions with no remaining ambiguity. 

\section{Hartree-Fock Computation of Ground State at Integer Fillings} \label{appendix_HF}
To evaluate the diamagnetic response we first perform a self-consistent
Hartree Fock calculation to obtain the closest Slater determinant
approximant to the ground state for a given set of parameters. For
this, consider the double-gated screened Coulomb interaction, 
\[
H_{V}=\frac{1}{2A}\sum_{\boldsymbol{q}}V_{\boldsymbol{q}}:\rho_{\boldsymbol{q}}\rho_{-\boldsymbol{q}}:,
\]
$A$ being the area of the sample, $V_{\boldsymbol{q}}=V_{0}d\,\tanh\left(qd\right)/q$
with $V_{0}$ defined in the main text, and the density operator given by 
\[
\rho_{\boldsymbol{q}}=\sum_{\boldsymbol{k}}\Lambda_{\alpha\beta}\left(\boldsymbol{k},\boldsymbol{k}-\boldsymbol{q}\right)c_{\boldsymbol{k},\alpha}^{\dagger}c_{\boldsymbol{k}-\boldsymbol{q},\beta}^{\phantom{\dagger}}.
\]
Here $\alpha,\beta$ are spin-band-valley indices and the projectors
are diagonal in valley due to the emergent $U(1)$ valley conservation and independent of spin. The form factors in the main text $\lambda_{\mu}^{\alpha,\beta}$, have the explicit diagonal valley index and no spin indices. Here we keep the most general matrix structure for simplicity: $\Lambda_{\alpha,\beta}(\vec{k},\vec{k}+\vec{q}) =\langle u_{\alpha}(\vec{k}) | u_{\beta}(\vec{k}+\vec{q}) \rangle$  with $|u_{\alpha}(\vec{k})\rangle$, the cell periodic wavefunctions obtained after the diagonalization of the BM Hamiltonian.
At this point we have not projected the interaction to the flat bands,
we can do so within mean field which generates a contribution to the
dispersion from the remote bands. Doing this, the Hamiltonian projected
to the flat bands takes the following form
\begin{align*}
H & =[H_{0}+ H_{V}]_{\rm active}\\
 & =\sum_{\boldsymbol{k}}c_{\boldsymbol{k},\alpha}^{\dagger} \left[ \tilde{h}_{0} \left(\vec{k}\right) \right]_{\alpha \beta}c_{\boldsymbol{k},\beta}^{\phantom{\dagger}}+\frac{1}{2A}\sum_{\boldsymbol{q}}V_{\boldsymbol{q}}\Lambda_{\alpha\beta}\left(\boldsymbol{k},\boldsymbol{k}-\boldsymbol{q}\right)\Lambda_{\alpha'\beta'}\left(\boldsymbol{k}',\boldsymbol{k}'+\boldsymbol{q}\right)c_{\boldsymbol{k},\alpha}^{\dagger}c_{\boldsymbol{k}',\alpha'}^{\dagger}c_{\boldsymbol{k}'+\boldsymbol{q},\beta'}^{\phantom{\dagger}}c_{\boldsymbol{k}-\boldsymbol{q},\beta}^{\phantom{\dagger}},
\end{align*}
where the $\alpha, \beta$ indices only take values in the active subspace. At this stage we change basis from the band to the sublattice basis. Furthermore, the renormalized Hamiltonian takes the form
\begin{equation}
\tilde{h}_{0}\left(\vec{k}\right)=h_{BM}\left(\vec{k}\right)+h_{\text{HF}}\left[P_{R}\right],  
\label{eqn:almost_bare_ham}
\end{equation}
where $P_{R}$ is the density matrix with the valence remote bands filled completely for all momenta, and $H_{\text{HF}}$
is the mean-field decomposition of the interaction. For generality,
we write the most general Hartree and Fock terms for a density matrix
that could break any flavor or translation symmetries, $\langle c_{\boldsymbol{k},\alpha}^{\dagger}c_{\boldsymbol{k}',\beta}^{\phantom{\dagger}}\rangle=P_{\alpha\beta}\left(\boldsymbol{k},\boldsymbol{k}'\right):$
\[
h_{\text{HF}}\left[P\right]\left(\boldsymbol{k},\boldsymbol{k}'\right)=h_{\text{H}}\left[P\right]\left(\boldsymbol{k},\boldsymbol{k}'\right)+h_{\text{F}}\left[P\right]\left(\boldsymbol{k},\boldsymbol{k}'\right).
\]
The explicit form of the Hartree and Fock terms are given by,
\[
h_{\rm H}^{\alpha \beta}\left[P\right]\left(\boldsymbol{k},\boldsymbol{k}'\right)=\sum_{G}V\left(\boldsymbol{k}'-\boldsymbol{k}+\boldsymbol{G}\right)M\left(\boldsymbol{k}'-\boldsymbol{k}+\boldsymbol{G}\right)\Lambda_{\alpha \beta} \left(\boldsymbol{k},\boldsymbol{k}'+\boldsymbol{G}\right),
\]
with
\[
M\left(\boldsymbol{k}'-\boldsymbol{k}+\boldsymbol{G}\right)=\frac{1}{A}\sum_{\alpha' \beta' \boldsymbol{p}}\Lambda_{\alpha'\beta'}\left(\boldsymbol{p}+\boldsymbol{k}',\boldsymbol{p}+\boldsymbol{k}-\boldsymbol{G}\right)P_{\alpha'\beta'}\left(\boldsymbol{k}'+\boldsymbol{p},\boldsymbol{k}+\boldsymbol{p}\right),
\]
and
\[
h_{\rm F}^{\alpha \beta}\left[P\right]\left(\boldsymbol{k},\boldsymbol{k}'\right)=-\frac{1}{A}\sum_{\boldsymbol{qG}}\sum_{\alpha'\beta'}V\left(\boldsymbol{q}-\boldsymbol{G}\right)\Lambda_{\alpha \alpha'}^{\dagger}\left(\boldsymbol{k}+\boldsymbol{q},\boldsymbol{k}+\boldsymbol{G}\right)P_{\beta',\alpha'}\left(\boldsymbol{k}'+\boldsymbol{q},\boldsymbol{k}+\boldsymbol{q}\right)\Lambda^{\phantom \dagger}_{\beta'\beta}\left(\boldsymbol{k}'+\boldsymbol{q},\boldsymbol{k}'+\boldsymbol{G}\right).
\]

We replace $h_{BM}$ in Eqn.~\ref{eqn:almost_bare_ham} with a `bare' Hamiltonian that does not include the renormalization of the parameters to avoid double counting \cite{bultinck_ground_2020}. This Hamiltonian defined by the relation
\begin{equation}
    h_{BM}(\vec{k}) = \tilde{h}_{BM}(\vec{k}) + \langle H_V \rangle_{P_0},
\end{equation}
 with $P_0$ being a reference density matrix. There are multiple prescriptions to select $P_0$, with all of them yielding qualitatively similar results in the calculation of the Hartree-Fock ground state under strain as reported in Ref.~\cite{kwan_kekule_2021}. For simplicity, we take the so-called `average' scheme in which $ P_0=\frac{1}{2}\mathbb{I}_{A} + P_R$ with $\mathbb{I}_{A}$ defined as the identity in the active degrees of freedom. Overall we make the replacement $h_{BM}(k) \rightarrow \tilde{h}_{BM}(k)$ in Eqn.~\ref{eqn:almost_bare_ham}. Since $P_0$ is a quadratic density matrix, we can readily evaluate the expectation value using Wick's theorem, which leads to the renormalized projected kinetic term,
\begin{equation}
\tilde{h}_{0}\left(\boldsymbol{k}\right)\rightarrow h^{*}_{0}\left(\boldsymbol{k}\right) = h_{BM}\left(\boldsymbol{k}\right)+h_{\text{HF}}\left[P_{R} - P_0 \right].  
\label{eqn:bare_ham}
\end{equation}
Overall the Hamiltonian that we use to carry out the self consistent determination of the ground state is 
\begin{equation}
H   =\sum_{\boldsymbol{k}}c_{\boldsymbol{k},\alpha}^{\dagger} \left[ h^*_{0} \left(\vec{k}\right) \right]_{\alpha \beta}c_{\boldsymbol{k},\beta}^{\phantom{\dagger}}+\frac{1}{2A}\sum_{\boldsymbol{q}}V_{\boldsymbol{q}}\Lambda_{\alpha\beta}\left(\boldsymbol{k},\boldsymbol{k}-\boldsymbol{q}\right)\Lambda_{\alpha'\beta'}\left(\boldsymbol{k}',\boldsymbol{k}'+\boldsymbol{q}\right)c_{\boldsymbol{k},\alpha}^{\dagger}c_{\boldsymbol{k}',\alpha'}^{\dagger}c_{\boldsymbol{k}'+\boldsymbol{q},\beta'}^{\phantom{\dagger}}c_{\boldsymbol{k}-\boldsymbol{q},\beta}^{\phantom{\dagger}}.
\end{equation}

We can connect to the discussion from the main text expressing the Hamiltonian in terms of $H_{\rm kin}$ and $H_{\rm int}$ by undoing the normal ordering and adding/subtracting the background charge density at neutrality. The former will yield a Fock contribution and the latter will yield a Hartree contribution. Overall the kinetic part of the Hamiltonian can then be written as 
\begin{equation}
    H_{\rm kin} = \sum_{\boldsymbol{k}}c_{\boldsymbol{k},\alpha}^{\dagger} \left[ h_{0} \left(\boldsymbol{k}\right)\right]_{\alpha \beta}c^{\phantom\dagger}_{\boldsymbol{k},\beta},
\end{equation}
with the matrix elements $h_{0}$ given by
\begin{align*}
 h_{0}\left(\boldsymbol{k}\right) & = h_{BM}\left(\boldsymbol{k}\right)+h_{\text{HF}}\left[P_{R} - P_0 + \frac{1}{2}\mathbb{I}_A\right]  \\
 & = h_{BM}\left(\boldsymbol{k}\right)
 \end{align*}
The fact that after all the renormalizations the kinetic part reduces to the usual BM model is a feature of the `average' subtraction scheme \cite{hofmann2021fermionic}. After normal ordering and subtracting the background charge density, the interaction Hamiltonian then takes the form
\begin{equation}
    H_{\rm int} = \frac{1}{2A}\sum_{\boldsymbol{q}}V_{\boldsymbol{q}}\delta\rho_{\boldsymbol{q}}\delta\rho_{-\boldsymbol{q}},
\end{equation}
with the additional definitions,
\begin{align}
    \delta\rho_{\boldsymbol{q}} &= \rho_{\boldsymbol{q}} -  \overline{\rho}_{\boldsymbol{q}},\\
     \overline{\rho}_{\boldsymbol{q}} &= \frac{1}{2}\sum_{\vec{k}\vec{G}}\delta_{\vec{q},\vec{G}}\mathrm{tr}\Lambda(\vec{k}, \vec{k}+\vec{G}),
\end{align}
which recovers then both $H_{\rm kin}$ and $H_{\rm int}$ in the main text.

 We now perform the Hartree-Fock decomposition in the bilinear:
  \begin{equation}
     \langle c^\dagger_{\vec{k}',\tau',m',s'} c^{\phantom\dagger}_{\vec{k},\tau, m,s} \rangle  = P_{m'\tau's',m\tau s}\left(\k',\k\right), 
 \end{equation}
 where the particular form of the mean field ansatz is given by:
 \begin{equation}
     P_{m'\tau's',m\tau s}\left(\k',\k\right) =\sum_{\p}\delta_{\p,\k'+\tau'\vec{Q}/2}\delta_{\p,\k+\tau \vec{Q}/2}\mathcal{P}_{m'\tau's',m\tau s}\left(\p\right),
     \label{eqn:ansatz_tsb}
 \end{equation}
with the vector $\vec{Q}$ left unspecified at the moment and we make explicit the valley, spin and sublattice indices, $\tau, s,$ and $m$, respectively. In practice we take the solution that minimizes the Hartree-Fock energy across all $\vec{Q}$ in showing our results in the main text. When $\Q=\vec{0}$ we recover the translation invariant solution possibly with flavor symmetry breaking in \cite{bultinck_ground_2020}. Using the ansatz above, the Hartree and Fock terms now become: 
\begin{equation}
    H_{\rm H} =\sum_{nn's\tau\tau'k}c_{n,\tau,s}^{\dagger}\left(\k-\tau \Q/2\right)\mathcal{H}_{\rm H}^{nn'\tau\tau'ss'}\left(\k-\tau \Q/2,\k-\tau'\Q/2\right)c^{\phantom\dagger}_{n',\tau',s'}\left(\k-\tau'\Q/2\right),
\end{equation}
where the matrix elements of the Hartree term are given by 
\begin{align}
    \mathcal{H}_{\rm H}^{nn'\tau\tau'ss'}\left(\k-\tau \Q/2,k-\tau'\Q/2\right) &=\frac{1}{A}\sum_{\G}V\left(\G\right) M\left(\G\right)\Lambda_{nn'}^{\tau}\left(\k-\tau \Q/2,\k-\tau'\Q/2+\G\right)\delta_{\tau\tau'}\delta_{ss'}\\
     M\left(\G\right) &= \sum_{s'\tau'mm'\p}\Lambda_{mm'}^{\tau'}\left(\p-\tau'\Q/2,\p-\tau'\Q/2-\G\right)\mathcal{P}_{m\tau's',m'\tau's'}\left(\p\right).
     \label{eqn:hart_shift}
\end{align}
Note that since the Hartree term is diagonal in valley, all of the dependence on $\Q$ can be removed by shifting the sums. However, we keep the dependence on $\k - \tau \Q/2$ to make manifest the emergent translation symmetry accompanied by a valley rotation for all the terms in the Hamiltonian. On the other hand, the Fock term is given by
\begin{equation}
    H_{\rm F} =\sum_{nn's\tau\tau'k}c_{n,\tau,s}^{\dagger}\left(\k-\tau \Q/2\right)\mathcal{H}_{F}^{nn'\tau\tau'ss'}\left(\k-\tau \Q/2,\k-\tau'\Q/2\right)c^{\phantom\dagger}_{n',\tau',s'}\left(\k-\tau'\Q/2\right),
\end{equation}
where the matrix elements of the Fock term are given by 
\begin{align}
    \mathcal{H}_{F}^{nn'\tau\tau'ss'}\left(\k-\tau \Q/2,\k-\tau'\Q/2\right)&=-\frac{1}{A}\sum_{\k'\G}\sum_{m'm}V\left(\k'-\k-\G\right)\Lambda_{nm}^{\tau\dagger}\left(\vec{k}'-\tau \Q/2,\vec{k}-\tau \Q/2+\G\right) \nonumber\\ & \times \mathcal{P}_{m'\tau's',m\tau s}\left(\vec{k}'\right)\Lambda_{m'n'}^{\tau'}\left(\vec{k}'-\tau'\Q/2,\vec{k}-\tau'\Q/2+\G\right).
    \label{eqn:fock_shift}
\end{align}

Finally, we shift the sum over momentum in the renormalized two-fermion term:
\begin{equation}
    H_{0}=\sum_{nn's\tau\tau'\k}c_{n,\tau,s}^{\dagger}\left(\k-\tau \Q/2\right)\mathcal{H}_{0}^{nn'\tau\tau'ss'}\left(\k-\tau \Q/2\right)c^{\phantom\dagger}_{n',\tau',s'}\left(\k-\tau'\Q/2\right),
\end{equation}
with the matrix elements given by 
\begin{equation}
    \mathcal{H}_{0}^{nn'\tau\tau'ss'}\left(\k-\tau \Q/2\right)=\left[h^{*}_{0}\left(\k-\tau \Q/2\right)\right]_{nn',\tau}\delta_{\tau\tau'}\delta_{ss'}.
\end{equation}

With all the explicit shifts over momentum, we can express the energy in terms of the matrix $\mathcal{P}$ only upon taking the expectation value. Adding the condensation energy, the total Hartree-Fock energy can then be expressed as 

\begin{equation}
    E_{\rm HF} = \sum_{\k}\mathrm{tr} \left( \mathcal{P}^T(\k)\left[ \mathcal{H}_0(\tilde{\k}) +\frac{1}{2}\mathcal{H}_{\rm HF}[\mathcal{P}](\tilde{\k}) \right] \right),
    \label{eqn:HFener}
\end{equation}
where we abbreviated the momentum dependence of the matrix elements of the Hamiltonian by the boosted momenta $\tilde{\k} = \k - \tau_z \Q/2$ and $\mathcal{H}_{\rm HF}$ is the sum of Eqn.~\ref{eqn:hart_shift} and Eqn.~\ref{eqn:fock_shift}. To obtain a self-consistent solution for $\mathcal{P}$ we use the optimal damping algorithm \cite{ODAalgo}. We run the algorithm in parallel for 32 initial random seeds and for all available $\Q$ in the $\k$-point mesh and select the solution that minimizes Eqn.~\ref{eqn:HFener}. Representative solutions for the band structure at high and low strain at different integer fillings are shown in Fig.\ref{fig:bands}. The bandwidth, $W^*$, that we report in Fig.2e is precisely the bandwidth of the set of active bands calculated self consistently.

\begin{figure*}[pth!]
\centering
\includegraphics[width=1.0\linewidth]{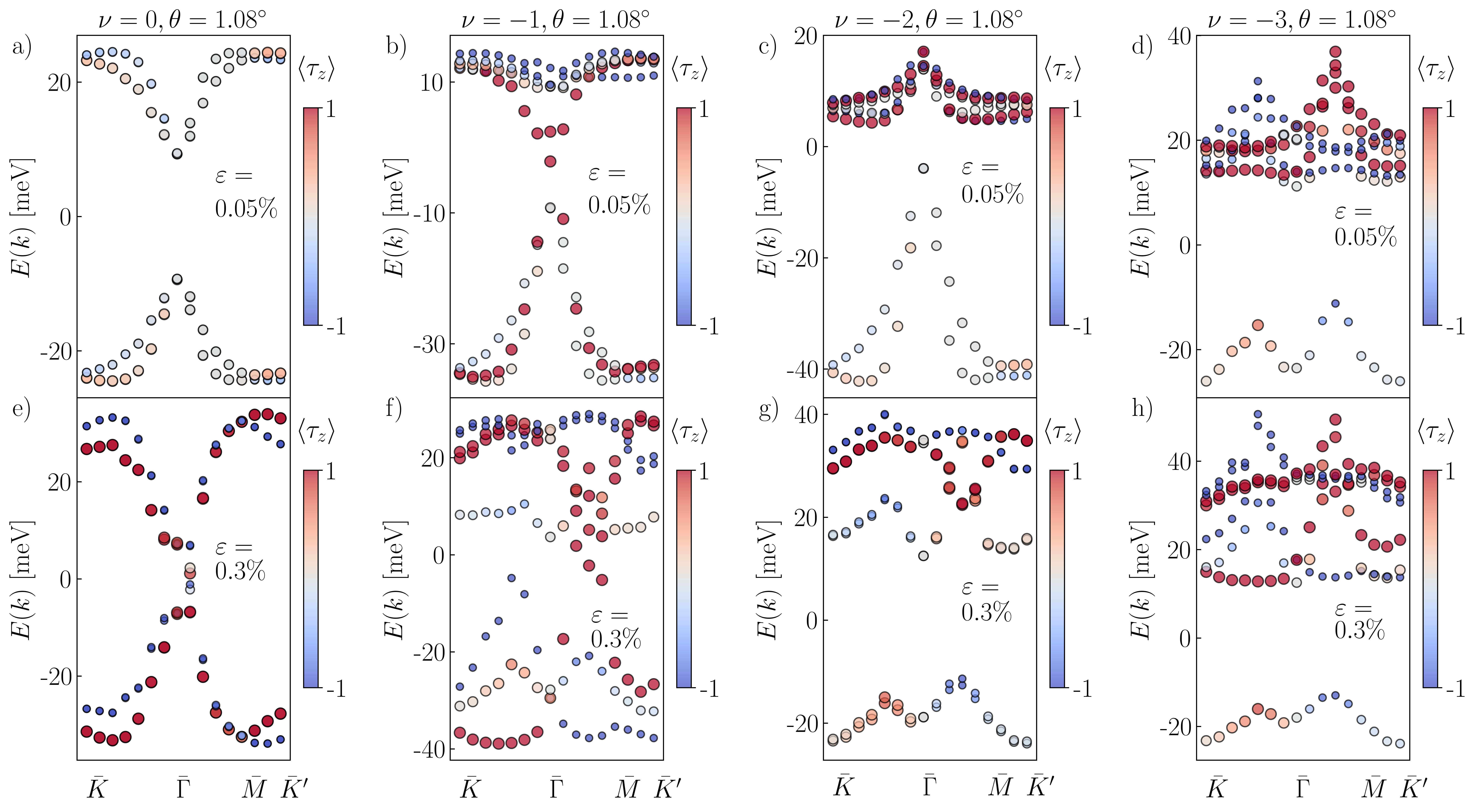}
\caption{Renormalized bandstructures obtained within Hartree-Fock computations at different integer fillings for a (a)-(d) strain $\varepsilon=0.05\%$, and (e)-(h) strain $\varepsilon=0.3\%$, respectively.
}
\label{fig:bands}
\end{figure*}
 
Numerically, we often need the wavefunctions at $\k\pm\Q/2$ or $\k\pm\Q/2\pm\G$, instead of recalculating the wavefunctions each time, we diagonalize the BM model over a fixed momentum grid in the first Brillouin zone and whenever we need the form factors at $\k+\G$ we use the periodic property:
\begin{equation}
u_{\tau,n,\G}(\k+\G')=u_{\tau,n,\G-\G'}(\k),
\label{eqn:periodic}
\end{equation}
where we explicitly write all the band/sublattice, valley and plane-wave indices in the BM Hamiltonian. In practice we have a finite plane-wave cutoff so that Eqn.~\ref{eqn:periodic} needs to be supplemented with an additional prescription for the edges of the $\G$ momentum lattice. In our case, we simply consider periodic boundary conditions. With this, we can express the wavefunctions at $\k\pm\Q/2\pm\G$ by rearranging indices in the original k-point grid and using Eqn.~\ref{eqn:periodic} to match the momenta as appropriate.

\section{Numerical evaluation of $K_{xx}$} \label{appendix_KXX}
Having obtained the estimate for the ground state density matrix, we can now proceed to evaluate the diamagnetic response. Here, we briefly review the formalism \cite{DMDC1,DMDC2} and apply it to the case of twisted bilayer graphene in the projected limit. When taking the entire spectrum into account, the current and the diamagnetic response can be obtained by performing a Peierls' substitution $t\left(\vec{r}-\vec{r}'\right)\rightarrow t\left(\vec{r}-\vec{r}'\right)e^{i\vec{A}\cdot\left(\vec{r}-\vec{r}'\right)}$ and taking the appropriate number of derivatives with respect to an external vector potential $\vec{A}$ in the limit where $\vec{A}\rightarrow0$. The current and the diamagnetic response are explicitly given by,
\begin{align}
    J_{\mu}\left(q_{\mu}\rightarrow0\right)	&=-\left.\frac{\delta H}{\delta A_{\mu}}\right|_{\vec{A}\rightarrow0} =-i\left[\hat{\mathcal{X}}_{\mu},H\right]\\
    K_{\mu\nu}&=\left.\frac{1}{2}\frac{\delta^{2}H}{\delta A_{\mu}\delta A_{\nu}}\right|_{\vec{A}\rightarrow0}=-\left[\hat{\mathcal{X}}_{\mu},\left[\hat{\mathcal{X}}_{\nu},H\right]\right],
\end{align}
where $\hat{\mathcal{X}}_{\mu}=\sum_{i}x_{i}^{\mu}c_{i}^{\dagger}c_{i}^{\phantom{\dagger}}$ is the many-body position operator. In the low-energy limit of a Hamiltonian that features a set of isolated flat bands for which the interaction $U$ is smaller than the gap to the remote bands $\Delta$, one can systematically integrate out the high-energy degrees of freedom via a Schriffer-Wolff transformation. The result of this procedure can be simply stated in terms of the replacement $H\left[\vec{A}\right]\rightarrow H_{\text{eff}}\left[\vec{A}\right]$ in the derivatives above. Here, $H_{\text{eff}}\left[\vec{A}\right]$ corresponds to the effective Hamiltonian where high-energy degrees of freedom have been effectively integrated out and the limit $U/\Delta\rightarrow0$ has been taken while keeping the probe vector potential. Taking the derivatives explictly, the effective low-energy operators are given by 
\begin{align}
    J_{\text{x}}^{\text{eff}}&=-i\left[\mathbb{P}\hat{\mathcal{X}}_{x}\mathbb{P},\mathbb{P}H\mathbb{P}\right],\\
    K_{\text{xx}}^{\text{eff}}&=-\left[\mathbb{P}\hat{\mathcal{X}}_{x}\mathbb{P},\left[\mathbb{P}\hat{\mathcal{X}}_{x}\mathbb{P},\mathbb{P}H\mathbb{P}\right]\right],
\end{align}
where $\mathbb{P}$ is the projector onto the low-energy subspace that is comprised by the set of isolated flat bands. The above result can be stated in terms of a projected gauge transformation of the Hamiltonian which emerges at low energies due to the effective number conservation in the limit $U/\Delta\rightarrow0$ :
\begin{equation}
    K_{\text{xx}}^{\text{eff}}=\frac{1}{A}\partial_{\alpha}^{2}\left.\left(e^{i\frac{e}{\hbar}\alpha\mathbb{P}\hat{\mathcal{X}}\mathbb{P}}\mathbb{P}H\mathbb{P}e^{-i\frac{e}{\hbar}\alpha\mathbb{P}\hat{\mathcal{X}}\mathbb{P}}\right)\right|_{\alpha=0},
\end{equation}
where we reinstated the fundamental constants and the appropriate dependence on the system size of $K_{\text{xx}}^{\text{eff}}$. The relation above can be verified by expanding the exponentials. We can carry out this calculation by performing the projected gauge transformation onto the creation and annihilation operators, which to leading order transform as:
\begin{align}
    e^{i\alpha\mathbb{P}\hat{\mathcal{X}}\mathbb{P}}c_{\boldsymbol{k}m\tau s}^{\phantom\dagger}e^{-i\alpha\mathbb{P}\hat{\mathcal{X}}\mathbb{P}} &=\sum_{m'\in\text{act}}c^{\phantom\dagger}_{\boldsymbol{k}+\alpha\boldsymbol{e}_{x}m'\tau s}\left\langle u^{\tau}_{\boldsymbol{k}m}|u^{\tau}_{\boldsymbol{k}+\alpha\boldsymbol{e}_{x}m'}\right\rangle +\frac{1}{2}\alpha^{2}g_{mm'}^{\tau, xx}\left(\boldsymbol{k}\right)c_{\boldsymbol{k}m'\tau s}^{\phantom\dagger},\\
   e^{i\alpha\mathbb{P}\hat{\mathcal{X}}\mathbb{P}}c_{\boldsymbol{k}n\tau s }^{\dagger}e^{-i\alpha\mathbb{P}\hat{\mathcal{X}}\mathbb{P}} &=\sum_{n'\in\text{act}}c_{\boldsymbol{k}+\alpha\boldsymbol{e}_{x}n'\tau s}^{\dagger}\left\langle u^{\tau}_{\boldsymbol{k}+\alpha\boldsymbol{e}_{x}n'}|u^{\tau}_{\boldsymbol{k}n}\right\rangle +\frac{1}{2}\alpha^{2}g_{n'n}^{\tau, xx}\left(\boldsymbol{k}\right)c_{\boldsymbol{k}n'\tau s}^{\dagger}.
\end{align}
where the explicit low-energy $U(1)$ valley conservation and the absence of spin-orbit coupling prevent the mixing of different valley and spin degrees of freedom by the projected gauge transformation, and $g_{n'n}^{\tau xx}\left(\boldsymbol{k}\right)$ is the multiorbital quantum metric defined by 
\begin{equation}
    g_{n'n}^{\tau,xx} = -\frac{1}{2} \langle u^{\tau}_{\k n'}| \partial_{k_x}^2 \Pi^{\tau}_{\k} |u^{\tau}_{\k n}  \rangle 
\end{equation}
with
\begin{equation}
    \Pi^{\tau}_{\k} = \sum_{n\in\text{act}} |u^{\tau}_{\k n}  \rangle \langle u^{\tau}_{\k n}|.
\end{equation}

With these definitions, we can then proceed with the calculation of $\kxx$ by applying the projected gauge transformations to the Hamiltonian in the set of active flat bands. Using $H = H_{\rm kin} + H_{\rm int}$ with the definitions from the previous section, we can distinguish between two types of terms in the Hamiltian, that contain two fermion operators and four fermion operators, respectively. We now treat these two types of terms separately. 

Starting with the two fermion terms, these can be grouped together in the form \begin{equation}
    H_{\text{2-ferm}}=\sum_{\tau s nm\boldsymbol{k}}\epsilon^{\tau}_{nm}\left(\boldsymbol{k}\right)c_{\boldsymbol{k}n\tau s}^{\dagger}c_{\boldsymbol{k}m \tau s},
    \label{eqn:2ferm_term}
\end{equation}
where $\epsilon^{\tau}_{nm}$ can be interpreted as a renormalized dispersion by the Hartree term originating from the subtraction of the charge density with respect to neutrality. More concretely
\begin{equation}
    \epsilon^{\tau}_{nm}(\k) = \left[ h_0 (\k)  \right]^{\tau}_{nm} - \frac{1}{A}\sum_{\G} V(\G) M(\G) \Lambda^{\tau}_{nm} (\k, \k + \G).
    \label{eqn:2fermdisp_for_kxx}
\end{equation}

Performing the projected gauge transformation on Eq.~\ref{eqn:2ferm_term} yields the following expression for the two-fermion contribution to $\kxx$
\begin{equation}
    \left.\kxx\right|_{\rm 2-ferm} = \left( \frac{e}{\hbar} \right)^2 \frac{1}{A}\sum_{n m\tau s \boldsymbol{k}} c_{\boldsymbol{k}\tau ns}^{\dagger}c^{\phantom \dagger}_{\boldsymbol{k}\tau m s}  \left[ \langle u_{\boldsymbol{k}n}^{\tau}|\frac{\partial^{2}\mathbb{E}^{\tau}}{\partial k_x ^{2}}|u_{\boldsymbol{k}m}^{\tau}\rangle + g_{nn'}^{\tau,xx}\left(\boldsymbol{k}\right)\epsilon_{n'm}\left(\boldsymbol{k}\right)+\epsilon_{nm'}\left(\boldsymbol{k}\right)g_{m'm}^{\tau,xx}\left(\boldsymbol{k}\right) \right]
\end{equation}
where we abbreviated the first term by definition
\begin{equation}
    \mathbb{E}^{\tau}\left(\boldsymbol{k}\right)=\sum_{nm}|u^{\tau}_{\boldsymbol{k}n}\rangle\epsilon^{\tau}_{nm}\left(\boldsymbol{k}\right)\langle u^{\tau}_{\boldsymbol{k}m}|.
\end{equation}
Now we proceed with the four-fermion term that originates from $H_{\rm int}$. Note that in the theory that includes all bands the only explicit dependence to $K_{\rm xx}$ would come from the kinetic term, however, in the projected theory, the external gauge field can couple to the wavefunction overlaps in the interaction giving a finite contribution. Performing the gauge transformation in the four fermion term, yields the following contribution
\begin{equation}
   \left.\kxx\right|_{\rm 4-ferm} =  \frac{1}{2} \left(\frac{e}{\hbar}\right)^{2}\frac{1}{A^2} \sum_{\tau\tau'}\sum_{ss'}\sum_{\boldsymbol{q}\boldsymbol{p}\boldsymbol{k}}\sum_{n_{1}n_{2}m_{1}m_{2}}\mathcal{U}_{n_{1}n_{2};m_{1}m_{2}}^{\tau\tau'}\left(\boldsymbol{k},\boldsymbol{p},\boldsymbol{q}\right)c_{\boldsymbol{k}\tau sn_{1}}^{\dagger}c_{\boldsymbol{k}-\boldsymbol{q}\tau sn_{2}}c_{\boldsymbol{p}\tau's'm_{1}}^{\dagger}c_{\boldsymbol{p}+\boldsymbol{q}\tau's'm_{2}}
   \label{eqn:4ferm_term}
\end{equation}

with the definition 
\begin{align}
    \mathcal{U}_{n_{1}n_{2};m_{1}m_{2}}^{\tau\tau'}\left(\boldsymbol{k},\boldsymbol{p},\boldsymbol{q}\right)&=\sum_{\boldsymbol{G}}V\left(\boldsymbol{q}+\boldsymbol{G}\right)\left\langle u_{\boldsymbol{k}n_{1}}^{\tau}|\partial_{k_{x}}^{2}\left(\Pi_{\boldsymbol{k}}^{\tau}\Pi_{\boldsymbol{k}-\boldsymbol{q}-\boldsymbol{G}}^{\tau}\right)|u_{\boldsymbol{k}-\boldsymbol{q}-\boldsymbol{G}n_{2}}^{\tau}\right\rangle \left\langle u_{\boldsymbol{p}m_{1}}^{\tau'}|u_{\boldsymbol{p}+\boldsymbol{q}+\boldsymbol{G}m_{2}}^{\tau'}\right\rangle \nonumber \\&+2\sum_{\boldsymbol{G}}V\left(\boldsymbol{q}+\boldsymbol{G}\right)\left\langle u_{\boldsymbol{k}n_{1}}^{\tau}|\partial_{k_{x}}\left(\Pi_{\boldsymbol{k}}^{\tau}\Pi_{\boldsymbol{k}-\boldsymbol{q}-\boldsymbol{G}}^{\tau}\right)|u_{\boldsymbol{k}-\boldsymbol{q}-\boldsymbol{G}n_{2}}^{\tau}\right\rangle \left\langle u_{\boldsymbol{p}m_{1}}^{\tau'}|\partial_{p_{x}}\left(\Pi_{\boldsymbol{p}}^{\tau'}\Pi_{\boldsymbol{p}+\boldsymbol{q}+\boldsymbol{G}}^{\tau'}\right)|u_{\boldsymbol{p}+\boldsymbol{q}+\boldsymbol{G}m_{2}}^{\tau'}\right\rangle \nonumber \\&+\sum_{\boldsymbol{G}}V\left(\boldsymbol{q}+\boldsymbol{G}\right)\left\langle u_{\boldsymbol{k}n_{1}}^{\tau}|u_{\boldsymbol{k}-\boldsymbol{q}-\boldsymbol{G}n_{2}}^{\tau}\right\rangle \left\langle u_{\boldsymbol{p}m_{1}}^{\tau'}|\partial_{p_{x}}^{2}\left(\Pi_{\boldsymbol{p}}^{\tau'}\Pi_{\boldsymbol{p}+\boldsymbol{q}+\boldsymbol{G}}^{\tau'}\right)|u_{\boldsymbol{p}+\boldsymbol{q}+\boldsymbol{G}m_{2}}^{\tau'}\right\rangle \nonumber \\&-\frac{1}{2}\sum_{\boldsymbol{G}}V\left(\boldsymbol{q}+\boldsymbol{G}\right)\left\langle u_{\boldsymbol{k}n_{1}}^{\tau}|u_{\boldsymbol{k}-\boldsymbol{q}-\boldsymbol{G}n_{2}}^{\tau}\right\rangle \left\langle u_{\boldsymbol{p}m_{1}}^{\tau'}|\left\{ \left[\partial_{p_{x}}^{2}\Pi_{\boldsymbol{p}}^{\tau'}\right]\Pi_{\boldsymbol{p}}^{\tau'}+\Pi_{\boldsymbol{p}+\boldsymbol{q}+\boldsymbol{G}}^{\tau'}\left[\partial_{p_{x}}^{2}\Pi_{\boldsymbol{p}+\boldsymbol{q}+\boldsymbol{G}}^{\tau'}\right]\right\} |u_{\boldsymbol{p}+\boldsymbol{q}+\boldsymbol{G}m_{2}}^{\tau'}\right\rangle \nonumber \\&-\frac{1}{2}\sum_{\boldsymbol{G}}V\left(\boldsymbol{q}+\boldsymbol{G}\right)\left\langle u_{\boldsymbol{k}n_{1}}^{\tau}|\left\{ \left[\partial_{k_{x}}^{2}\Pi_{\boldsymbol{k}}^{\tau}\right]\Pi_{\boldsymbol{k}}^{\tau}+\Pi_{\boldsymbol{k}-\boldsymbol{q}-\boldsymbol{G}}^{\tau}\left[\partial_{k_{x}}^{2}\Pi_{\boldsymbol{k}-\boldsymbol{q}-\boldsymbol{G}}^{\tau}\right]\right\} |u_{\boldsymbol{k}-\boldsymbol{q}-\boldsymbol{G}n_{2}}^{\tau}\right\rangle \left\langle u_{\boldsymbol{p}m_{1}}^{\tau'}|u_{\boldsymbol{p}+\boldsymbol{q}+\boldsymbol{G}m_{2}}^{\tau'}\right\rangle 
    \label{eqn:horrible_expr}
\end{align}

Together, Eqn.~\ref{eqn:2ferm_term} and Eqn.~\ref{eqn:4ferm_term} form the basis of our calculation of $\langle \kxx \rangle$ in Fig.2a. Since we have a finite plane wave cutoff, we have to take a finite number of terms in the sum over $\G$ processes. We verified that results converged for our set of parameters restricting the sum up to two Umklapp processes. 

We can now use the states from Hartree Fock to evaluate the expectation value of $\kxx$, using Wick's theorem, Using the particular translation symmetry breaking ansatz in Eqn.~\ref{eqn:ansatz_tsb}, we get

\begin{align}
    \left\langle \left.\kxx\right|_{\rm 4-ferm} \right\rangle &=\frac{1}{2}\left(\frac{e}{\hbar}\right)^{2}\frac{1}{A^{2}}\sum_{\tau\tau'}\sum_{ss'}\sum_{\boldsymbol{p}\boldsymbol{k}}\sum_{n_{1}n_{2}m_{1}m_{2}}\left[\mathcal{U}_{n_{1}n_{2};m_{1}m_{2}}^{\tau\tau'}\left(\boldsymbol{k}-\tau\boldsymbol{Q}/2,\boldsymbol{p}-\tau'\boldsymbol{Q}/2,\boldsymbol{0}\right)\mathcal{P}_{n_{1}\tau sn_{2}\tau s}\left(\boldsymbol{k}\right)\mathcal{P}_{m_{1}\tau's'm_{2}\tau's'}\left(\boldsymbol{p}\right)\right.\nonumber\\&+\left.\mathcal{U}_{n_{1}n_{2};m_{1}m_{2}}^{\tau\tau'}\left(\boldsymbol{k}-\tau\boldsymbol{Q}/2,\boldsymbol{p}-\tau'\boldsymbol{Q}/2,\boldsymbol{k}-\boldsymbol{p}\right)\left(\delta_{\tau'\tau}\delta_{s's}\delta_{m_{1}n_{2}}-\mathcal{P}_{m_{1}\tau's'n_{2}\tau s}\left(\boldsymbol{p}\right)\right)\mathcal{P}_{n_{1}\tau sm_{2}\tau's'}\left(\boldsymbol{k}\right)\right]
    \label{eqn:4-ferm_expt}
\end{align}

For the two-fermion piece, we find 
\begin{align}
    \left\langle  \left.\kxx\right|_{\rm 2-ferm} \right\rangle= \left( \frac{e}{\hbar} \right)^2 \frac{1}{A}\sum_{n m\tau s \boldsymbol{k}} \mathcal{P}_{n \tau s m\tau s} (\k - \tau \Q /2) \left[ \langle u_{\boldsymbol{k}n}^{\tau}|\frac{\partial^{2}\mathbb{E}^{\tau}}{\partial k_x ^{2}}|u_{\boldsymbol{k}m}^{\tau}\rangle + g_{nn'}^{\tau,xx}\left(\boldsymbol{k}\right)\epsilon_{n'm}\left(\boldsymbol{k}\right)+\epsilon_{nm'}\left(\boldsymbol{k}\right)g_{m'm}^{\tau,xx}\left(\boldsymbol{k}\right) \right]
    \label{eqn:2-ferm_expt}
\end{align}

The value of $\langle\kxx \rangle_{\rm int}$ in Fig2a corresponds to the sum of  $\left\langle \left.\kxx\right|_{\rm 4-ferm} \right\rangle$ (Eqn.~\ref{eqn:4-ferm_expt}) and the Hartree contribution in $\left\langle  \left.\kxx\right|_{\rm 2-ferm} \right\rangle$ (Eqn.~\ref{eqn:2-ferm_expt}) while $\langle\kxx \rangle_{\rm kin}$ corresponds to the contribution from the bare kinetic energy to $\left\langle  \left.\kxx\right|_{\rm 2-ferm} \right\rangle$.

\section{$\kxx$ at other integer fillings}

In this section, we present results (Figs.~\ref{fig:neut}, \ref{fig:m1} and \ref{fig:m3}) for the optical spectral weight at other integer fillings, as we go away from the chiral flat-band limit (where the response vanishes), including the effects of strain. We focus on neutrality ($\nu=0$), where increasing strain stabilizes a semi-metallic phase, and the odd integer fillings ($\nu=-1,~-3$), where the system has a strong tendency towards developing a QAH phase (with additional momentum-dependent structure). 

We find that, overall, the spectral weight at large $\varepsilon$ decreases away from charge neutrality. A larger spectral weight does not necessarily imply stronger tendency towards superconductivity; near neutrality the reduced density of states as one dopes away from the semimetallic state potentially interferes with superconductivity. For odd-integer fillings, the appearance of time-reversal symmetry breaking due to the spin polarization in the Kekul\'e spirals at large $\varepsilon$ can induce pair-breaking effects if the superconducting state has a large component overlapping with spin-singlet pairing.

\begin{figure*}[pth!]
\centering
\includegraphics[width=1.0\linewidth]{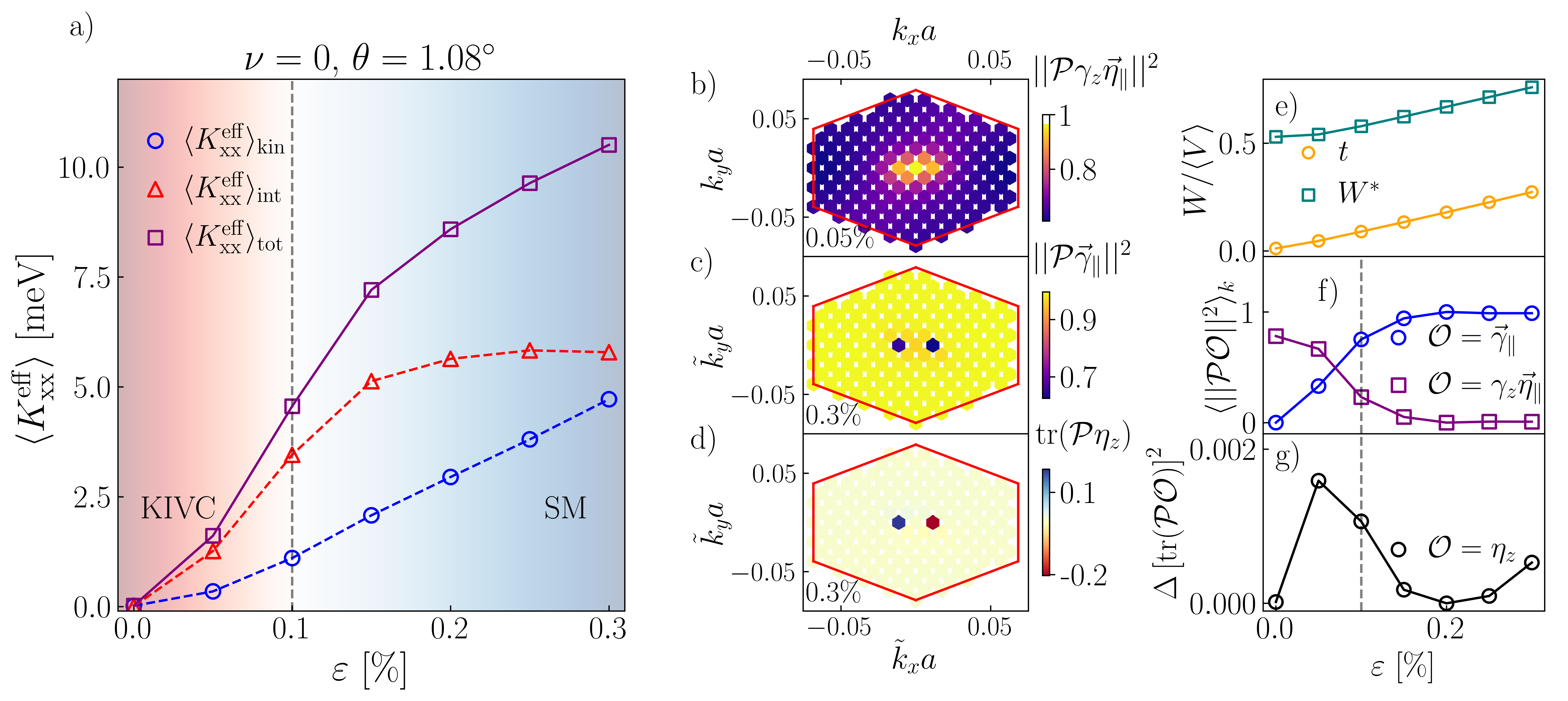}
\caption{(a) The effective diamagnetic response (see Eq.~\ref{eq:partial_f_sum}) evaluated at $T=0$ and for $\nu = 0$ (instead of $\nu=-2$ as in Fig.~\ref{fig:approx_kxx}) as a function of strain within a fully self-consistent Hartree-Fock computation. The dashed line demarcates the transition from a KIVC to a semimetal (SM). (b)-(d) Order parameters characterizing the ground state for $\varepsilon=0.05\%$ and $\varepsilon=0.3\%$, respectively. In (c) and (d), we use the momentum-boosted mini-Brillouin zone.To account for the gauge dependence of $\vec{\gamma}_{\parallel}=(\gamma_x,\gamma_y)$ and the phase of the IVC order parameter, we plot $||\mathcal{PO} ||^2 =|\mathrm{tr}_{\gamma \eta}(\mathcal{PO})^2|$ in (b), (c), where $\mathrm{tr}_{\gamma \eta}$ is the partial trace over valley and Chern pseudospin and $|\cdot|$ is the Frobenius norm.  (e) Evolution of $t/\langle V\rangle$ and $W^*/\langle V\rangle$, where $W^*$ represents the interaction induced bandwidth with increasing strain. (f) The momentum-averaged order parameters in (b) and (c) with increasing $\varepsilon$. 
(g) Variation of the momentum-averaged fluctuation in the valley polarization as a function of strain. 
}
\label{fig:neut}
\end{figure*}

\begin{figure*}[pth!]
\centering
\includegraphics[width=1.0\linewidth]{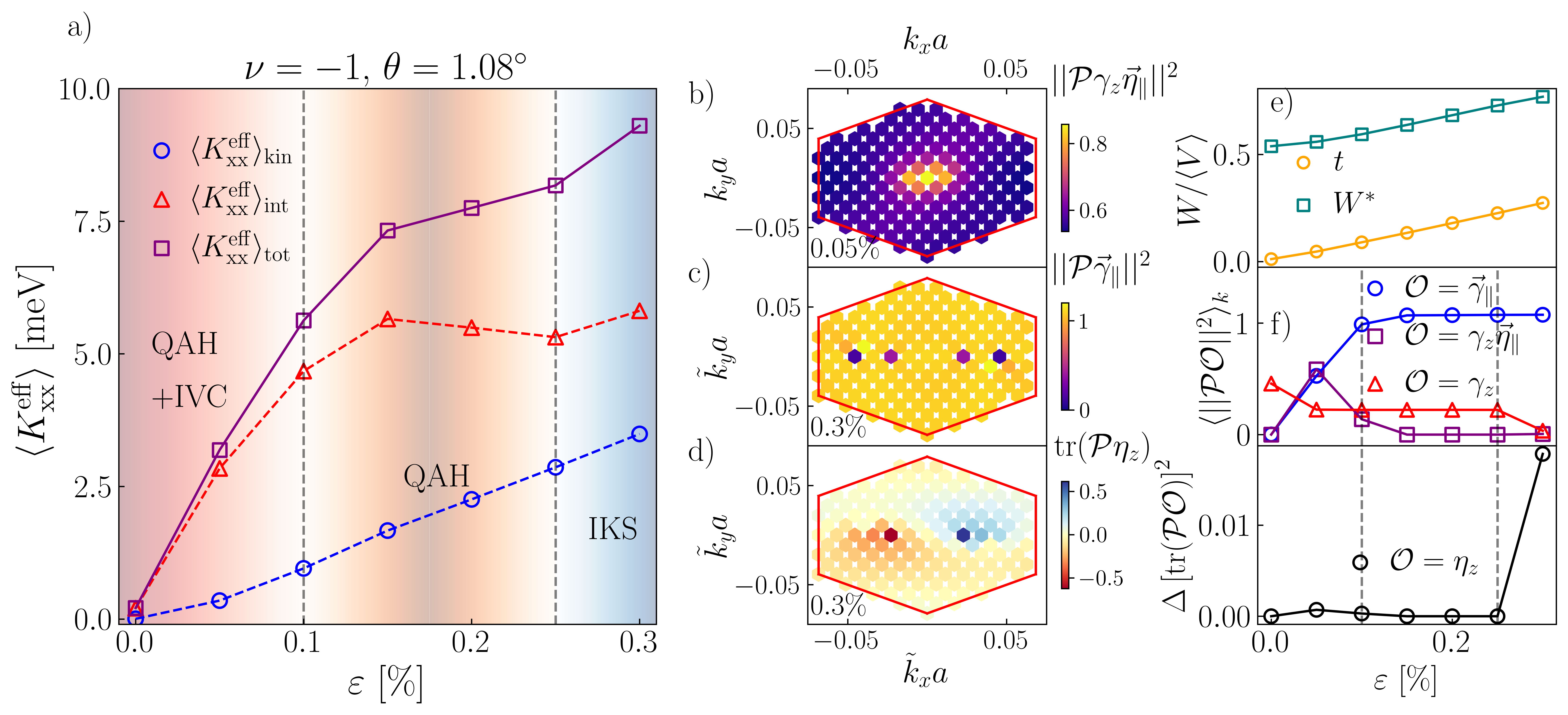}
\caption{Results for the effective diamagnetic response at $T=0$ for $\nu=-1$. The individual quantities being plotted are same as in Figs.~\ref{fig:approx_kxx} and \ref{fig:neut}. The noteworthy difference at the odd-integer filling is the appearance of the QAH phases over a broad range of low to intermediate strains.   
}
\label{fig:m1}
\end{figure*}

\begin{figure*}[pth!]
\centering
\includegraphics[width=1.0\linewidth]{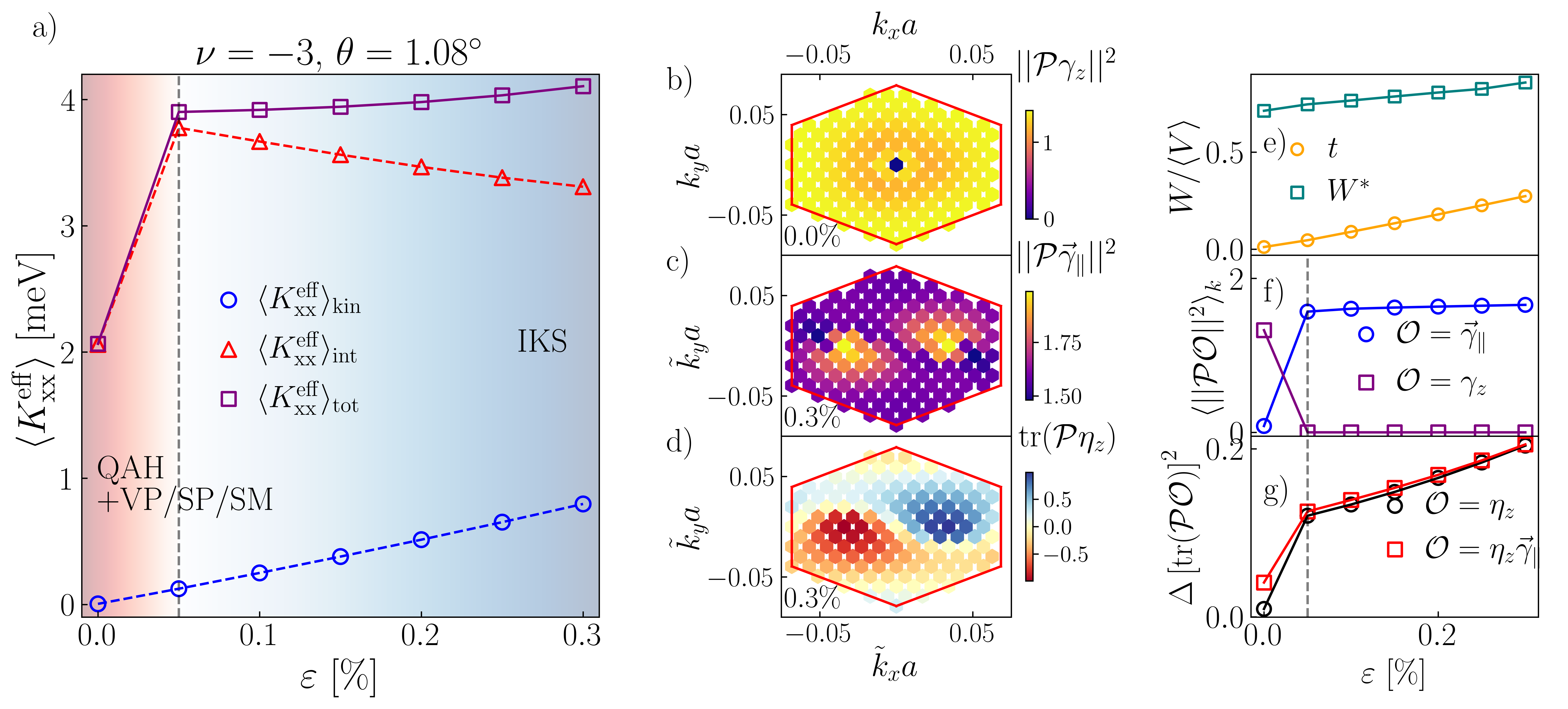}
\caption{Results for the effective diamagnetic response at $T=0$ as in Fig.~\ref{fig:m1} but for $\nu=-3$. The non-vanishing $\langle \kxx \rangle$ at zero strain is related to the fact that the state has a momentum-dependent component along $\eta_z \vec{\gamma}_{\parallel}$.
}
\label{fig:m3}
\end{figure*}

\section{Twist angle dependence}
We have studied the dependence of the effective diamagnetic response on twist-angle at two different integer fillings and for two different values of the strain in Fig.~\ref{fig:theta}. At small strain, the predominant non-monotonic variation is controlled by the contribution from the kinetic energy, while the interaction contribution is monotonic. At large values of the strain and in the absence of pure KIVC order, the variations become much smaller and the response becomes very weakly dependent on the twist-angle. 
\begin{figure*}[pth!]
\centering
\includegraphics[width=1.0\linewidth]{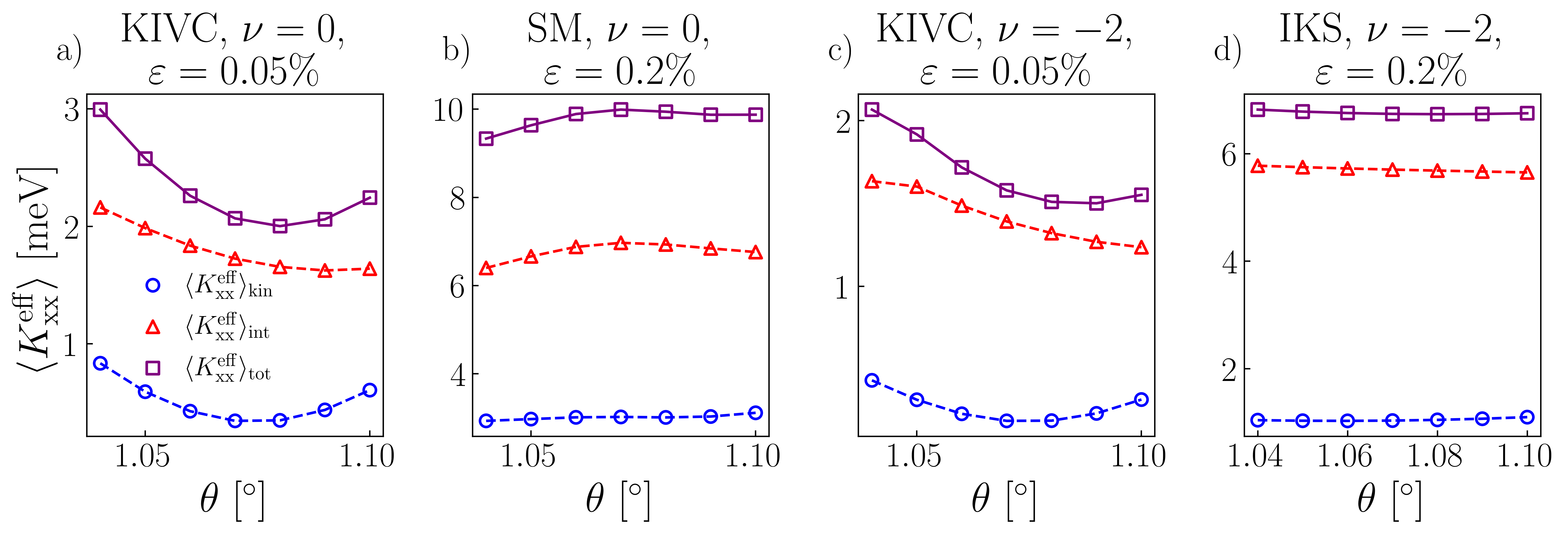}
\caption{ Results for the effective diamagnetic response at $T=0$ evaluated as a function of twist-angle ($\theta$) at (a)-(b) $\nu=0$, and (c)-(d) $\nu=-2$ at two different values of the strain. 
}
\label{fig:theta}
\end{figure*}

\end{widetext}

\end{document}